\documentclass[twoside]{article}
\usepackage{qic}
\newcommand{\beq}[0]{\begin{equation}}
\newcommand{\eeq}[0]{\end{equation}}
\newcommand{\bw}[0]{\begin{widetext}}
\newcommand{\ew}[0]{\end{widetext}}
\newcommand{\bc}[0]{\begin{center}}
\newcommand{\ec}[0]{\end{center}}
\newcommand{\bwn}[0]{\begin{widetext}\begin{eqnarray}}
\newcommand{\ewn}[0]{\end{eqnarray}\end{widetext}}
\newcommand{\beqn}[0]{\begin{eqnarray}}
\newcommand{\eeqn}[0]{\end{eqnarray}}

\newcommand{\tzn}[0]{{\it i.e.}}

\newcommand{\uroj}[0]{\mathrm{i}}
\newcommand{\eksp}[0]{\mathrm{e}}
\newcommand{\proj}[1]{|#1\rangle \langle #1|}
\newcommand{\ket}[1]{|#1\rangle}
\newcommand{\bra}[1]{\langle #1 |}
\newcommand{\dager}[1]{{#1}^{\dagger}}

\newcommand{\outerp}[2]{\ket{#1}\! \bra{#2}}
\newcommand{\inner}[2]{ \langle #1 | #2 \rangle}

\newcommand{\linia}[0]{\newline\noindent}

\newcommand{\kan}[0]{\Lambda}
\newcommand{\oper}[0]{\calL(\varrho)}
\newcommand{\tr}[0]{\mathrm{tr}}

\newcommand{\jedynka}[0]{\mathbbm{1}}

\newcommand{\non}[0]{\nonumber\\}

\newcommand{\spann}[0]{\mathrm{span}}

\newcommand{\pruf}[0]{\noindent \textbf{Proof:}\;}
\newcommand{\prufend}[0]{$\blacksquare$}
\newtheorem{property}[theorem]{Property}
\newtheorem{observation}[theorem]{Observation}
\newtheorem{fact}[theorem]{Fact}
\newcommand{\ods}[0]{\vspace*{12pt}
\noindent
}

\def\calB{{\cal B}}
\def\calC{{\cal C}}
\def\calD{{\cal D}}

\def\calH{{\cal H}}

\def\calL{{\cal L}}

\def\calP{{\cal P}}
\def\calQ{{\cal Q}}
\def\calR{{\cal R}}
\def\calS{{\cal S}}
\def\calT{{\cal T}}
\def\calU{{\cal U}}

\def\frakS{{\frak S}}

\usepackage{amssymb,amsmath}
\usepackage{graphicx}
\usepackage{dcolumn}
\usepackage{bm}
\usepackage{bbm}

\begin{document}
\setlength{\textheight}{8.0truein}    

\runninghead{Title  $\ldots$}
            {Author(s) $\ldots$}

\normalsize\textlineskip
\thispagestyle{empty}
\setcounter{page}{1}

\copyrightheading{0}{0}{2003}{000--000}

\vspace*{0.88truein}

\alphfootnote

\fpage{1}

\centerline{\bf
MULTIACCESS QUANTUM COMMUNICATION}
\vspace*{0.035truein}
\centerline{\bf AND PRODUCT HIGHER RANK NUMERICAL RANGE }
\vspace*{0.37truein}
\centerline{\footnotesize
MACIEJ DEMIANOWICZ}
\vspace*{0.015truein}
\begin{center}
\footnotesize\it Atomic Physics Division, Department of Atomic Physics and Luminescence, \\Faculty of Applied Physics and Mathematics,\\
Gda\'nsk University of Technology, ul. Narutowicza 11/12, PL80-233 Gda\'nsk, Poland \\ National Quantum Information Center in Gda\'nsk, ul. W\l. Andersa 27,
PL81-824 Sopot, Poland \\ maciej@mif.pg.gda.pl\end{center}
\baselineskip=10pt
\vspace*{10pt}
\centerline{\footnotesize
PAWE\L\; HORODECKI}
\vspace*{0.015truein}
\begin{center}
\footnotesize\it Department of Theoretical Physics and Quantum Informatics, Faculty of Applied Physics and Mathematics,\\
Gda\'nsk University of Technology, ul. Narutowicza 11/12, PL80-233 Gda\'nsk, Poland  \\ National Quantum Information Center in Gda\'nsk, ul. W\l. Andersa 27,
PL81-824 Sopot, Poland\\ pawel@mif.pg.gda.pl\end{center}
\vspace*{10pt}
\centerline{\footnotesize
KAROL \.ZYCZKOWSKI}
\vspace*{0.015truein}
\begin{center}
\footnotesize\it Smoluchowski Institute of Physics, Jagiellonian University, ul. Reymonta 4, 30-059 Krak\'ow, Poland \\ Center of Theoretical Physics, Polish Academy of Science, al. Lotnik\'ow 32/46, 02-668 Warszawa, Poland \\ karol@cft.edu.pl\end{center}
\vspace*{0.225truein}

\vspace*{0.21truein}

\abstracts{
In the present paper we initiate the study of the product higher rank numerical range. The latter, being a variant of the higher rank numerical range [M.--D. Choi {\it et al.}, Rep. Math. Phys. {\bf 58}, 77 (2006); Lin. Alg. Appl. {\bf 418}, 828 (2006)],
is a natural tool for studying a construction of quantum error correction codes for multiple access channels.
We review properties of this set and relate it to other numerical ranges, which were recently introduced in the literature.
Further, the concept is applied to the construction of codes for
bi--unitary two--access channels with a hermitian noise model. Analytical techniques for both outerbounding the product higher rank numerical range and determining its exact shape are developed for this case. Finally,
the reverse problem of constructing a noise model for a given product range is considered.
}{}{}

\vspace*{10pt}

\keywords{multiparty quantum channel, product higher rank numerical range, quantum error correction, Knill--Laflamme conditions}
\vspace*{3pt}
\communicate{to be filled by the Editorial}

\vspace*{1pt}\textlineskip    
\section{Introduction}
Quantum information transmission \cite{bennett-huge,lloyd,barnum-nielsen-schumacher,BKN,devetak} inevitably involves occurrence of errors which faced not properly can disable faithful communication. Developing fruitful and useful strategies for combating these errors is thus one of the main challenges of the theory of quantum channels. Many effort has been put in this direction
and several techniques have been developed to overcome destructive influence of coupling to the environment (see, {\it e.g.}, \cite{byrde-et-al,kribs-et-al}).
Among them, quantum error correction codes (QECC) along with the celebrated Knill--Laflamme (KL) conditions \cite{KL} are the most widely recognized (see \cite{gaitan-ksiazka} and references therein).
Methods of constructing QECC for quantum communication have been previously reported in the literature \cite{steane,calderbank-shor,gottesman}. There has also been a  significant progress in experimental realizations of the propositions \cite{cory-et-al,leung-et-al,knill-exp}.

Recently, a fruitful approach to solving Knill--Laflamme conditions has been put forward \cite{choi-et-al,choi-et-al2,majgier}. It relies on the concept of the higher rank numerical range of an operator and provides a systematic framework for a construction of QECC. As shown in Ref. \cite{choi-et-al,choi-et-al2}, it may capture more possibilities than the stabilizer formalism \cite{gottesman}.

 However, the proposals analyzed so far concerned
bipartite communication --- no general approach has been developed to treat the case of the larger number of users of a quantum network (see however \cite{wilde}  and the related paper \cite{demianowicz-2012}). The main purpose of this paper is to provide a mathematical tool of the {\it product higher rank numerical range} for designing QECC for multiaccess quantum communication. Product higher rank numerical range by the definition is a higher rank numerical range restricted to product projections.

The paper is organized as follows. First, we review basic ideas, including error correction, of quantum communication over (multiparty) quantum channels. Further, we briefly recall some basic features of the higher rank numerical range with applications to error correction. We then move to the main body of the present paper by introducing the notion of the product higher rank numerical range and discussing its basic properties. In the next section we consider construction of QECC for bi--unitary channels with a hermitian noise model. We then demonstrate exemplary applications of our findings to some concrete problems. We also discuss the reverse problem of constructing a noise model for which a product code exists. The manuscript is concluded with a discussion.

\section{Quantum channels and quantum error correction}
Here we briefly recall some basic ideas of communication over quantum channels and set the scenario for further considerations.
\subsection{Quantum channels}
Quantum channel $\calL$ is a completely positive trace--preserving map. Every channel admits the so--called Kraus (or operator--sum) representation as follows $\calL(\varrho)=\sum_i A_i \varrho A_i^{\dag}$ with $\sum_i A_i^{\dag}A_i=\jedynka$ \cite{Choi,Kraus}. A random unitary channel is the one which has the representation $\calL(\varrho)=\sum_i p_i U_i \varrho U_i^{\dag}$, where $U_i$ are unitary and $\sum_i p_i =1$, $p_i\ge 0$. When such a channel has two Kraus operators, \tzn, $\calL(\varrho)=pU_1\varrho\dager{U_1}+(1-p)U_2\varrho\dager{U_2}$, it is called a bi--unitary channel (BUC). This kind of channels is the main interest of the present paper.

Channels can be classified according to the number of senders and receivers using them. We have the  following types of channels according to such a classification \cite{BKN,md-ph-1,yard,md-ph-2,broadcast}:
(i) bipartite --- one sender and a single receiver,
(ii) multiple access channels (MACs)--- several senders and one receiver,
(iii) broadcast --- one sender and several receivers,
(iv) $km$--user --- $k$ senders transmit information to $m$ receivers ($k,m>1$).

In our reasonings we mainly concentrate on two--access channels, that is multiple access channels with two senders.

Due to the possibility of a global rotation $U_i^\dag (\cdot)U_i$
on the output or the input of a channel, in the bipartite,
multiple access, and broadcast case one can consider a simplified
BUC in general reasonings\footnote{This is also true for a general
$km$--user channel if one of the unitaries $U_1$ or $U_2$ is
product across the cuts corresponding  to the separation of either
senders or receivers} \beq\label{buc}
\oper=p\varrho+(1-p)U\varrho\dager{U}. \eeq For two--access
channels it holds $\varrho=\varrho_1\otimes\varrho_2$, where
$\varrho_i$ is the input of the $i$--th sender.

\subsection{Quantum error correction}

QECC is a subspace $\calC$ of a larger Hilbert space $\calH$. Equivalently, a code is defined to be the projection $R_{\calC}$ onto $\calC\subseteq\calH$. One says that $\calC$ is correctable if all states from this subspace $\varrho=R_{\calC}\varrho R_{\calC}$ can be recovered after an action of a channel using some decoding operation $\calD$, that is $\calD\circ\calL (\varrho)=\varrho$. Such recovery operation exists if and only if
\beq \label{KL}
R_\calC  A_i^{\dag}A_j R_\calC=\alpha_{ij}R_\calC \eeq
for some hermitian matrix $[L]_{ij}=\alpha_{ij}$. These conditions are due to Knill and Laflamme (KL) \cite{KL}.

In the case of a larger number of senders we talk about local codes $\calC_i$, that is QECC for every sender.
It is an immediate observation that KL conditions need only a little adjustment to serve for the case of MACs. Namely, we have (with the obvious notation):
\begin{observation}
Local codes $\calC_i$ are correctable for a MAC with Kraus operators $\{A_i\}$ with $k$ inputs if and only if
\beq\label{KL-MAC} \left( R_{\calC_1}\otimes R_{\calC_2}\otimes\dots\otimes R_{\calC_k}\right)  A_i^{\dag}A_j \left(R_{\calC_1}\otimes R_{\calC_2}\otimes\dots\otimes R_{\calC_k}\right)=\alpha_{ij}R_{\calC_1}\otimes R_{\calC_2}\otimes\dots\otimes R_{\calC_k}
\eeq
 for some hermitian matrix $[\calL]_{ij}=\alpha_{ij}$.
\end{observation}
This is true since the set of product codes is a subset of the set of all codes. In further parts, we sometimes use the notation $R\otimes R'$ ($R_M\otimes R'_N$) or $S\otimes S'$ for a code for a two--access channel and talk about $M\otimes N$ codes, where $M,N$ denote dimensions of local codes.

In case of many usages of a channel, $A_i$ are tensor products of Kraus operators in KL conditions. In this paper, however, we concentrate on a single usage of a channel.
For one use of a BUC, Eq. (\ref{buc}), KL conditions (\ref{KL}) reduce to the {\it single} condition (we write $R_C$ shortly as $R$).
\beq\label{zwykly-single}
 RUR=\lambda R, \eeq
  which for MACs takes the form
\beq\label{condition-zwykly}
(R\otimes R')U(R\otimes R')=\lambda R\otimes R'.
\eeq

It is useful to introduce the notion of the entropy of a QECC \cite{kribs-pasieka-entropy-code}.
This entropy quantifies the number of ancillary qubits which are needed for the recovery procedure.
 For a BUC the entropy of a code is the von Neumann entropy of the matrix
 \beq\label{entropia-buc}
 \calL=\left(
         \begin{array}{cc}
           p & \lambda \sqrt{p(1-p)} \\
           \lambda^*\sqrt{p(1-p)} & 1-p \\
         \end{array}
       \right).
 \eeq
By inspection one finds that the entropy $S(\calL)=H(1/2(1+\sqrt{1-4(1-p)(1-|\lambda|^2)}))$. It is equal to zero iff
 $\lambda=\mathrm{e}^{\mathrm{i}\varphi}$ and these values correspond to so--called decoherence free subspaces (DFS) for which recovery is trivial (identity) recovery operation\footnote{It is not clear to us whether it would make any sense to define the entropy of a local code.} .
\subsection{Higher rank numerical range approach to bipartite QEC}\label{higher-approach}
Here we recall the notion of the higher rank numerical range\footnote{It is the generalization of the notion of the numerical range  of an operator $X$, which is defined
to be the following set: $\Lambda(X)=\{\lambda\in\mathbb{C}: \bra{\psi}X\ket{\psi}=\lambda,\langle\psi|\psi\rangle=1\}$}
{ }and its implication in the area of QEC
\cite{choi-et-al,choi-et-al2}.
The approach we briefly describe below is the one we wish to  modify further to be applicable in case of multiple access channels.

It is the form of KL conditions, which prompted the authors of
Ref. \cite{choi-et-al,choi-et-al2,choi-oam,gau-et-al} to introduce
the notion of the higher rank numerical range (or the rank--$k$
numerical range) of an operator. For an operator $A$, it is
defined to be the following set \beq
\Lambda_k(A)=\{\lambda\in\mathbb{C}: P_k A P_k=\lambda P_k\} \eeq
with $P_k \in \calP_k$, where $\calP_k$ is the set of rank $k$
projections. Elements of the set are sometimes called the
compression values.

 Full characterization of the set $\Lambda_k(A)$ for hermitian $A$ has been obtained. Namely, assuming $a_1\le a_2\le \ldots\le a_N$ is a spectrum of an $N\times N$ hermitian $A$, it holds that $\Lambda_k(A)=\langle a_k, a_{N-k+1} \rangle$, which is (a) a true interval whenever $a_k < a_{N-k+1}$, (b) a singleton set if $a_k = a_{N-k+1}$, (c) an empty set in the remaining case.

The set is also quite well understood for unitary operators \cite{choi-oam}. We recall some results below.
Let $U$ be an $n\times n$ unitary matrix with a non--degenerate spectrum $\mathrm{spec}(U)=\{{z_i}\}_{i=1}^n$ corresponding to eigenvectors $\{\ket{v_i}\}_{i=1}^n$.
Let $\Delta_k(U)$ be the set of $\lambda$ such that for some $k$ disjoint subsets $\delta_1,\delta_2,\dots,\delta_k$ of $\{1,2,\dots,n\}$
 it holds $\lambda\in conv(\{z_i,i\in \delta_j\})$ for all $i$ ($conv$ stands for the convex hull). It was shown\footnote{In fact for $n\ge 3k$ much stronger result was proved, namely $\Delta_k=\Lambda_k$ but to avoid technicalities we use here a weaker version, which is sufficient for our purposes.}{ }  that $\Delta_k(U)\subseteq\Lambda_k(U)$.
The proof of this fact is constructive in a sense that it gives explicitly the projection. Since
\beq\label{compress}
\lambda=\sum_{j\in \delta_i}\alpha_{ij}z_i
\eeq
 with $\alpha_{ij}\ge 0$ and $\sum_{j\in\delta_i}\alpha _{ij}=1$ we can choose the code $P_C$ to be
 $
 P_C=\sum_{i=1}^k \proj{\psi_i}
$
 where
 \beq\label{weks}
 \ket{\psi_i}=\sum_{j\in\delta_i} \sqrt{\alpha_{ij}}\ket{v_j}
 \eeq
  to obtain $P_CUP_C=\lambda P_C$. One can take the subsets $\delta_i$ to represent triangles or, in a more restricted variant, sections. Both cases will be considered by us in Section \ref{toy}.

Relevance of the notion to the issue of construction of QECC can be easily recognized if one compares the definition of the higher rank numerical range with the form of KL conditions \cite{choi-et-al,choi-et-al2}.

\section{Product higher rank numerical range and its basic properties}\label{prod-higher}
\noindent Motivated by the form of KL conditions, Eq.
(\ref{KL-MAC}), for multiple access channels we introduce the
notion of the {\it  product} higher rank numerical range. It is
defined as follows \ods
\begin{definition}
The $k_1\otimes k_2\otimes\cdots$ product higher rank numerical range of an operator $A$ is defined to be
\beq\label{product}
\Lambda_{k_1\otimes k_2\otimes\cdots}(A)=\{\lambda\in\mathbb{C}:\left(R\otimes R'\otimes\cdots \right)A
\left( R\otimes R'\otimes\cdots\right)=\lambda R\otimes R'\otimes\cdots\}
\eeq
where $R\in \calP_{k_1},\;R'\in\calP_{k_2},\ldots$
\end{definition}
\ods

In the above, we assume that  $k_i\ne 1$ for at least  a single
index $i$ . When all $k_i$ are equal to unity then one deals with
the local (or product) numerical range \cite{product-local}, that
is a set \beq \Lambda_{\mathrm{loc}}(A)=\{\lambda\in\mathbb{C}:
\bra{\psi\otimes\phi}A\ket{\psi\otimes\phi}\;\;\mathrm{for}\;\;
\ket{\psi}\in\calH_a,\;\ket{\phi}\in\calH_b\}. \eeq

If we choose all projections to be the same then we deal with the
{\it symmetric} product higher rank numerical range
$\kan^{\mathrm{symm.}}_{k\otimes k\otimes\cdots}(A)$. On the other
hand, if only some of  projections are the same we call the set
{\it locally symmetric} product higher rank numerical range
$\kan^{\mathrm{loc.\;symm.  }\vec{p}}_{k\otimes k\otimes k'\otimes
\cdots}(A)$, where $\vec{p}$
 specifies which projections are to be chosen the same.
 Obviously, we can demand projections to be equal only when they project on subspaces of spaces with equal dimensionality.

One has:
\ods
\begin{fact}
Let $A$ act on $\mathbb{C}^d\otimes\mathbb{C}^d$.
It holds
\beq
\kan_{m\otimes m} (A)=\bigcup_{U} \kan_{m\otimes m}^{\mathrm{symm.}} \left((\jedynka\otimes U) A (\jedynka \otimes U^{\dagger})\right)=
\bigcup_{V} \kan_{m\otimes m}^{\mathrm{symm.}} ((V\otimes\jedynka) A (V^{\dagger}\otimes\jedynka)).
\eeq
\end{fact}
\ods
\textbf{Proof:} The defining equation (\ref{product}) can be rewritten as $[R\otimes (U^{\dagger}RU)] A [R\otimes (U^{\dagger}RU)]=\lambda R\otimes (U^{\dagger}RU)$ for some unitary $U$. It is equivalent to $(R\otimes R) (\jedynka\otimes U) A (\jedynka\otimes U^{\dagger})(R\otimes R)=\lambda R\otimes R$ from which the result follows. The second case is shown in a similar manner. $\blacksquare$

\noindent Naturally, a similar fact holds in the multipartite case.

We propose to call the $k_1\otimes k_2\otimes\cdots$ product higher rank numerical range  {\it multipartite} as opposed to the {\it bipartite} one, which corresponds to the $k_1\otimes k_2$ case.

We also propose to use the dual set, the $k_1\otimes k_2\otimes \cdots$ {\it product codes set} for $A$, defined as follows
\beq
V_{k_1\otimes k_2\otimes\cdots}(A)=\{R\in\calP_{k_1},\; R'\in\calP_{k_2}, \dots: (R\otimes R'\otimes\cdots) A(R\otimes R'\otimes\cdots)=\lambda R\otimes R'\otimes\cdots\}.
\eeq
\indent In a standard, \tzn\; non--product case, there is more than one projection (it may even be an infinite number, when the spectrum of an operator is degenerated) corresponding to the same compression value. This degeneracy may be removed in a product case, but we do not know to what extent this happens in a generic case.

Although we will be interested mainly in BUCs for which KL
conditions give the single equation (\ref{condition-zwykly}) to
solve, it is natural to introduce the notion of the {\it joint}
product higher rank numerical range (just as it is defined in the
standard case \cite{joint}). Namely, one defines \ods
\begin{definition}
The $k_1\otimes k_2\otimes\cdots$ {\it joint} product higher rank numerical range of operators $A_i$, $i=1,2,\ldots,I$, is defined to be the following set
\beqn
&&\hspace{-18pt}\Lambda_{k_1\otimes k_2\otimes\cdots}^{\mathrm{joint}}(A_1,A_2,\ldots,A_I)=\non &&\{(\lambda_1,\lambda_2,\ldots,\lambda_I)\in\mathbb{C}^{I}:(R\otimes R'\otimes\cdots )A_i(R\otimes R'\otimes\cdots)=\lambda_i R\otimes R'\otimes\cdots\}
\eeqn
where $R\in \calP_{k_1},\;R'\in\calP_{k_2},\ldots$
\end{definition}

For further convenience, we also introduce the {\it common} higher
rank numerical range which is defined as follows \ods
\begin{definition}
The $k_1\otimes k_2\otimes\cdots$ {\it common} product higher rank numerical range of operators $A_i$, $i=1,2,\ldots,I$, is defined to be the following set
\beqn
\Lambda^{\mathrm{comm.}}_{k_1\otimes k_2\otimes\cdots}(A_1,A_2,\ldots,A_I)=\{\lambda\in\mathbb{C}:(R\otimes R'\otimes\cdots) A_i(R\otimes R'\otimes\cdots)=\lambda R\otimes R'\otimes\cdots\}
\eeqn
where $R\in \calP_{k_1},\;R'\in\calP_{k_2},\ldots$
\end{definition}

In the present paper, we will be interested solely in the cases when the product structure of the projectors corresponds to the tensor product structure of a Hilbert space (which is usually uniquely determined by the problem under consideration).

The most striking difference between the standard and the product higher rank numerical range is the fact that while the former is determined solely by the eigenvalues of an operator, the latter would also be affected by the form of eigenvectors (in applications to a construction of QECC it is enough to consider normal operators). This makes the product range difficult to determine even with the knowledge of the standard one.

Product higher rank numerical range bears the following natural features
\footnote{In what follows we use $\boxtimes$ to denote the Minkowski product of two sets
 on the complex plane, which is defined as follows: $Z_1 \boxtimes Z_2 = \{z : z = z_1z_2; z_1 \in Z_1; z_2 \in Z_2\}$.} :
\begin{property}
$\Lambda_{m\otimes n}\subseteq \Lambda_{mn}$.
\end{property}
\begin{property}  $\Lambda_{m\otimes n}$ can be empty even when $\Lambda_{mn}$ is non--empty.
\end{property}
\begin{property}
$\Lambda_{m\otimes n}\subseteq \Lambda_{\mathrm{loc}}$.
\end{property}
\begin{property}
$\Lambda_{m\otimes n}(A)$ and $V_{m\otimes n}(A)$ are both compact sets.
\end{property}
\begin{property}\label{zawieranie-w-mniejszych}
$\Lambda_{(m_1+m_2)\otimes n}(A)\subseteq \Lambda_{m_i\otimes n}(A)$, $i=1,2$.
\end{property}
\begin{property}
$\Lambda_{m\otimes n}(A\otimes B)=\kan_m(A)\boxtimes \kan_n(B)$.
\end{property}
\begin{property}
$\kan_{m\otimes n}(A)=\kan_{m\otimes n}(U\otimes V A U^{\dagger}\otimes V^{\dagger})$ for arbitrary unitary $U$ and $V$.
\end{property}
\begin{property}
$\kan_{m\otimes n}^{\mathrm{comm.}}(A,B)\subseteq \kan_{m\otimes n}^{\mathrm{joint}}(A,B)$.
\end{property}
\begin{property}
$\kan_{m\otimes m\otimes m}^{\mathrm{symm.}}(A)
 \subseteq \kan_{m\otimes m\otimes m}^{\mathrm{loc.\;symm. \vec{p}}}(A)\subseteq \kan_{m\otimes m\otimes m}(A)$.
\end{property}
 \ods

 Before we proceed, we also need to recall the notion of the $C$--numerical range $W_C(A)$ of an operator $A$, which goes as follows \cite{c-range}:
 \beq
 W_C(A)=\{\tr C^{\dagger} U^{\dagger} A U, U\in \calU\}.
 \eeq
 When $U$s are taken to be product one deals with the {\it local} $C$--numerical range $W_C^{\mathrm{loc.}} (A)$ \cite{c-range-local}. In cases considered in the present paper the set is a closed interval. With this notion in hand we can give the following bound on the product higher rank numerical range:
 \begin{observation}\label{prod-w-C}
It holds that
\beq
\kan_{k_1\otimes k_2\otimes\cdots}(A)\subseteq W^{\mathrm{loc.}}_{R\otimes R'\otimes\cdots}\displaystyle \left(\frac{1}{k_1 k_2\cdots} A \right).
\eeq
 \end{observation}
{\bf Proof:} It is enough to take trace of both sides of the defining equation (\ref{product}) of the product range to conclude that every
$\lambda$ belonging to $\kan_{k_1\otimes k_2\otimes\cdots}(A)$ must also belong to $W^{\mathrm{loc.}}_{R\otimes R'\otimes\cdots}\displaystyle \left(A/k_1 k_2\cdots\right)$. $\blacksquare$

Naturally, we assume that $W^{\mathrm{loc.}}$ is local according to the same cut as the higher rank range is product.

The above observation will turn out to be very useful in bounding the product higher rank numerical range. Its application, however, will require a numerical optimization.

 We single out also some chosen properties  of the common range, which will serve as a basis for one of the examples. 
  \ods
\begin{property}\label{wspolny-dopelnien}
$\Lambda_k^{\mathrm{comm.}}(A,\jedynka-A)=\{\frac{1}{2}\}$ or $\emptyset$.
\end{property}
\begin{property}\label{wspolny-projektorow} Let $Q=\sum_{i=1}^{l} \proj{i}\otimes Q_i$, with orthonormal basis $\{\ket{i}\}$ and rank $q_i$ projections $Q_i$,
act on $\mathbb{C}^{d_1}\otimes \mathbb{C}^{d_2}$. Then the
following hold: (i) if $l=d_1$ then $\Lambda_{d_1\otimes
k}(Q)=\Lambda^{\mathrm{comm.}}_{ k}(Q_1, Q_2,\ldots,Q_{d_1})$,
(ii) if $l<d_1$ then either $\Lambda_{d_1\otimes k}(Q)=\emptyset$
or $\Lambda_{d_1\otimes k}(Q)=\{0\}$ with the latter holding if
and only if $0\in \Lambda^{\mathrm{comm.}}_{k}(Q_1,
Q_2,\ldots,Q_{l})$.
\end{property}
\begin{property}\label{ograniczenia-na-common}
$\Lambda^{\mathrm{comm.}}_k(A_1,A_2, \cdots, A_K)\subseteq \Lambda_k(\sum_{i=1}^K \alpha_i A_i)$ with $\sum_i \alpha_i =1$. In particular, for $K=2$, $\Lambda^{\mathrm{comm.}}_k(A_1,A_2)\subseteq\Lambda_k(\alpha A_1+(1-\alpha)A_2)$.
\end{property}
\ods
\noindent\textbf{Proof of Property \ref{wspolny-dopelnien}:} Adding (i) $RAR=\lambda R$ and  (ii) $R(\jedynka-A)R=\lambda R$ we obtain $R=2\lambda R$, thus $\lambda=1/2$ if equation (i) has a solution with this value, if it does not --- $\Lambda_k(A)$ is empty and so is $\Lambda_k^{\mathrm{comm.}}(A,\jedynka-A)$. $\blacksquare$

\noindent\textbf{Proof of Property \ref{wspolny-projektorow}:} (i) We look for $\lambda$ obeying
\beq\label{podwojny}
(R\otimes R') \left(\sum_{i=1}^{d_1} \proj{i}\otimes Q_i\right)( R\otimes R')=\lambda R\otimes R'
\eeq
for projections $R$ and $R'$ rank, respectively, $d_1$ and $k$.
Since $R$ must be of full rank it must be that $R=\jedynka_{d_1}$
and it ultimately follows that
\beq
R' Q_i R'=\lambda R',\quad \forall_i,
\eeq
concluding the proof of this part. (ii) Replacing upper limit in the sum with $l<d_1$ and considering diagonal terms (again $R=\jedynka_{d_1}$) we obtain
\beq
R'Q_i R'=\lambda R', \quad i=1,2,\ldots,l
\eeq
and  $d_1-l$ equations $0=\lambda R'$. Thus $\lambda$ must be equal to zero and this value must be in the common range of $Q_i$, $i=1,2,\ldots,l$. The claim then follows. $\blacksquare$

\noindent\textbf{Proof of Property \ref{ograniczenia-na-common}:} Let $K=2$ for simplicity. We have (i) $RA_1R=\lambda R$ and (ii) $RA_2R=\lambda R$. We multiply the first equation by $\alpha$, the second by $1-\alpha$ and add such equations to obtain (iii) $R(\alpha A_1+(1-\alpha)A_2)=\lambda R$. Thus each $\lambda$ which fulfills (i-ii) for some $R$, fulfills also (iii). The result then follows. $\blacksquare$

Our special interest in finding product higher rank numerical ranges of projections stems from the type of noise we mainly focus on in the paper (see Eq. (\ref{postac-u})).

\section{Two--access quantum communication: QECC for a BUC}
We now specify the type of noise we will further consider. We
assume that $U$ is hermitian. Hermiticity of a unitary matrix
implies that it must be of the form \beq\label{postac-u} U=P-Q
\eeq with some projections $P$, $Q$, such that $P+Q=\jedynka$. In
this case $S(\calL)=0$ iff $\lambda=\pm 1$.
\subsection{Zero entropy codes($\lambda=\pm 1$)}
In Ref. \cite{demianowicz-2012} the problem of the existence of zero entropy codes (decoherence free subspaces) for a channel given by Eq. (\ref{buc}) with the noise model Eq. (\ref{postac-u}) was
 formulated as the problem of judging decomposability\footnote{Consider the following transformation on matrices $\calT: A\longrightarrow EAF$ with nonsingular $E$ and $F$. A space of $d\times d$ matrices is called $(i,j)$--decomposable if all elements of it can be simultaneously brought with $\calT$ to the form in which they have a block zero matrix of size $(d-i)\times  (d-j)$ in the same position.    Let $\calS$ be a subspace of $\calH\subseteq \mathbb{C}^d\otimes \mathbb{C}^d$ with a projection $S=\sum_k \proj{\phi_k}$, $\ket{\phi_k}=\sum_{ij}c_{k}^{ij}\ket{ij}$.
  The matrices $[C_k]_{ij}\equiv c_{k}^{ij}$  are called Schmidt matrices and constitute a basis for a space, say $\frakS$.
Subspace $\calS$ is called $(i,j)$--decomposable if the space $\frakS$ is so.}{ } of subspaces $\calP\equiv P \calH$ and $\calQ\equiv Q \calH$ and
the following theorem was proved

\ods
\begin{theorem}\label{glowne-tw}
A $M\otimes N$ DFS
exists if and only if at least one of the subspaces $\calP$ or $\calQ$ is  $(d-M,d-N)$--decomposable.
\end{theorem}
\ods
It was also shown that it can be generalized to the generic case of $U$ allowing for DFS, \tzn, $U=P_0+\sum_k \eksp^{\uroj \beta_k} Q_k$ with $r(P_0)\ge MN$.

In Ref. \cite{demianowicz-2012} we have applied the theorem to several concrete examples. We recall two of them (in both cases $\calQ=\spann\{\ket{\phi_i}\}$):
\begin{itemize}
\item  for the following choice of $\phi_i$ no $2\otimes 2$ code  exists
\beqn
  \ket{\phi_1}=\frac{1}{\sqrt{2}}(\ket{11}+\ket{22}),\nonumber\\
  \ket{\phi_2}=\frac{1}{\sqrt{2}}(\ket{10}+\ket{21}),\nonumber
  \eeqn
\item for the following choice a $2\otimes 2$ code does exist
\beqn
  \ket{\phi_1}=\frac{1}{\sqrt{2}}(\ket{02}+\ket{10}),\nonumber\\
  \ket{\phi_2}=\frac{1}{\sqrt{2}}(\ket{01}+\ket{20}).\nonumber
  \eeqn
\end{itemize}
\subsection{Higher entropy codes($\lambda\ne \pm 1$)}\label{higher}
\noindent

We now move to the case of higher entropy codes, that is for $\lambda \ne\pm 1$ in Eq. (\ref{condition-zwykly}).

Let us start with some general remarks concerning the non--product case, that is  Eq. (\ref{zwykly-single}). Assuming Eq. (\ref{postac-u}) and denoting
 \beq\label{gama}
 \gamma=(1-\lambda)/2
 \eeq
   we obtain
 \beq\label{podstawa}
 RQR=\gamma R,
 \eeq
or, equivalently,
 \beq\label{rownowazne}
 RPR=(1-\gamma)R.
  \eeq
  It is almost evident that if Eq. (\ref{podstawa}) is to be fulfilled the rank (in this context understood in a standard sense) of $Q$ must be at least as large as the rank of $R$ but it cannot be too large since Eq. (\ref{rownowazne}) must also hold.
 The theory of the higher rank numerical range \cite{choi-et-al,choi-et-al2} makes this intuition strict and we have that $0\le \gamma \le 1$ only when $r(R)\le r(Q)\le n-r(R)$, where $n$ is the dimension of the whole space. We will later prove a result of this kind for ranks of the reduced matrices of $P$ and $Q$. It is clear that in such situations if $\gamma\in \Lambda(Q)$ then it is also the case for $1-\gamma$ for some other projection. In this way we have a dual pair of projections, say $R$ and $\hat{R}$ (see also the upcoming Corollary \ref{wektory-postac}). Notice that both codes have exactly the same entropy (see Eq. (\ref{entropia-buc})).

The main result of this section is the necessary condition for $\gamma$ to belong to the product higher rank  numerical range of a projection.
\begin{theorem}\label{konieczny}
Let $Q_l$ be a rank $l$ projection. Assume $(R_M\otimes R_N' )Q_l (R_M\otimes R_N') = \gamma R_M\otimes R_N'$ holds.
Let further $x_1\ge x_2\ge\dots$ be eigenvalues of $\mathrm{tr}_B Q_l$, and $y_1\ge y_2\ge\dots$ of $\mathrm{tr}_A Q_l$. Then
\beqn\label{ograniczenie}
\sum_{i=1}^M |x_i-N\gamma|+\sum_{i=M+1}^{r(\mathrm{tr}_B Q_l)} x_i\le MN\sqrt{(1-\gamma)(1+3\gamma)}+l-MN,\\
\sum_{i=1}^N |y_i-M\gamma|+\sum_{i=N+1}^{r(\mathrm{tr}_A Q_l)} x_i\le MN\sqrt{(1-\gamma)(1+3\gamma)}+l-MN.\nonumber
\eeqn
\end{theorem}
\ods
The proof of the theorem is based on several lemmas (some concerning also the non--product case), which we shall prove below.\ods
 \begin{lemma}\label{postac}
 Fix a number $n$.
 Let $R=\sum_{i=1}^k \proj{e_i}$ be a projection operator (the states $\ket{e_i}$ are orthonormal) onto a subspace of an $n$ dimensional space $\calH$. Denote with $\calB _R=\{\ket{e_1},\ket{e_2},\dots,\ket{e_k}\}$ and $\calB^{\perp}$ its orthonormal complement, so that $\spann\{\calB _R\oplus\calB^{\perp}\}=\calH $. If Eq. (\ref{podstawa}) holds with $0<\gamma<1$ then $Q$ must have the following form in $\calB_R\oplus\calB^{\perp}$,
 \beq
 Q=\left[
\begin{array}{c|c|c}
\begin{array}{ccc}
  \gamma &  &    \\
   &  \ddots &  \\
   &  &
   \gamma
\end{array}
 & \begin{array}{ccc}
  \sqrt{\gamma(1-\gamma)} &  &   \\
   &  \ddots &  \\
   &   & \sqrt{\gamma(1-\gamma)}
\end{array} & 0\\ \hline
 \begin{array}{ccc}
  \sqrt{\gamma(1-\gamma)} &  &  \\
   &  \ddots &  \\
    &  & \sqrt{\gamma(1-\gamma)}
\end{array} & \begin{array}{ccc}
  1-\gamma &  &    \\
   &   \ddots &  \\
   &    & 1-\gamma
\end{array}  & 0\\ \hline
0 & 0 & S
\end{array}\right],
 \eeq
where nonzero blocks with off diagonal terms equal to zero are of size $k\times k$ and $S$ is a projector of size $(n-2k)\times (n-2k)$.
 \end{lemma}
 \ods
 \pruf If Eq. (\ref{podstawa}) holds then $Q$, which is hermitian, must necessarily have the form
 \beq\label{ku}
 Q=\left(
   \begin{array}{cc}
     \gamma \jedynka_k & A \\
     \dager{A} & B \\
   \end{array}
 \right)
 \eeq
 with $\jedynka_k$ in the basis of the eigenvalues of $R$. Here $A$ is a rectangular $k\times (n-k)$ matrix, while $B$ denotes a positive semidefinite square matrix of size $n-k$.
 Since $Q$ is a projection it must be true that $Q^2=Q$. For convenience, we explicitly write the left hand side of this equation
 \beq\label{kukwadrat}
 Q^2=\left(
       \begin{array}{cc}
         \gamma^2 \jedynka_k+A\dager{A} & \gamma A+AB  \\
         \gamma \dager{A}+B\dager{A} & \dager{A}A+B^2 \\
       \end{array}
     \right).
 \eeq
 Comparison of Eqs. (\ref{ku}) and (\ref{kukwadrat}) gives us the set of conditions
 \beq\label{pierwszy}
  \gamma^2 \jedynka_k+A\dager{A}= \gamma \jedynka_k,
  \eeq
  \beq\label{drugi}
  \dager{A}A+B^2=B,
 \eeq
 \beq\label{trzeci}
 \gamma A+AB=A.
 \eeq
 From the first one we obtain
 \beq
 A\dager{A}=x^2\jedynka_k,\quad x^2=\gamma-\gamma^2 > 0.
 \eeq
 Denoting $\tilde{A}=A/x$ we can rewrite this as $\tilde{A}\dager{\tilde{A}}=\jedynka_k$.  Thus $\tilde{A}\dager{\tilde{A}}$ must be rank $k$ as $r(\jedynka_k)=k$.
  Since for an arbitrary $\tilde{A}$ it holds that $r(\tilde{A}\dager{\tilde{A}})=r(\tilde{A})$ and $r(\tilde{A})\le\min\{k,n-k\}$ we infer that $n-k\ge k$ as otherwise we would get a contradiction. Moreover it follows that $\tilde{A}$ is an isometry so we have
 \beq
 A=x\tilde{A}=x\sum_{i=1}^k \outerp{e_i}{v_i}
 \eeq
 with some orthonormal $\ket{v_i}\in\mathbb{C}^{n-k}$ (here we have $k$ such states but this set can naturally be completed to have $n-k$ elements so that $\calH=\spann\{\ket{e_1},\ket{e_2},\dots,\ket{e_k},\ket{v_1},\ket{v_2},\dots,\ket{v_{n-k}}\}$).
 We thus have
 \beq\label{postac-AA}
 \dager{A}A=x^2\sum_{i=1}^k\proj{v_i}.
  \eeq
  From Eq. (\ref{drugi}) we conclude that $[\dager{A}A,B]=0$ since an operator commutes with its own function. Along with Eq. (\ref{postac-AA}) this implies that $B$ has the following form
  \beq
  B=\left(\sum_{i=1}^k \lambda_i \proj{v_i}\right) \oplus B'
  \eeq
  where $B'$ lives on the subspace spanned by some orthonormal set $\{\ket{v_i}\}_{i=k+1}^{n-k}$. Moreover, from Eq. (\ref{trzeci}) it follows that $\lambda_i\equiv 1-\gamma$. Thus, taking into account that the sign of $x$ corresponds just to a global phase for basis vectors and so we can choose $x>0$, the matrix $Q$
   in the basis $\spann\{\calB _R\oplus\calB^{\perp}\}$, where $\calB _R=\{\ket{e_1},\ket{e_2},\dots,\ket{e_k}\}$ and $\calB^{\perp}=\{\ket{v_1},\ket{v_2},\dots,\ket{v_{n-k}}\}$, reads
  \beq
  Q=\left[
\begin{array}{c|c|c}
\begin{array}{ccc}
  \gamma &  &    \\
   &  \ddots &  \\
   &  &
   \gamma
\end{array}
 & \begin{array}{ccc}
  \sqrt{\gamma(1-\gamma)} &  &   \\
   &  \ddots &  \\
   &   & \sqrt{\gamma(1-\gamma)}
\end{array} & 0\\ \hline
 \begin{array}{ccc}
  \sqrt{\gamma(1-\gamma)}  &  &  \\
   &  \ddots &  \\
    &  & \sqrt{\gamma(1-\gamma)}
\end{array} & \begin{array}{ccc}
  1-\gamma &  &    \\
   &   \ddots &  \\
   &    & 1-\gamma
\end{array}  & 0\\ \hline
0 & 0 & B'
\end{array}\right].
 \eeq
This concludes the proof.
 \prufend
\ods
\begin{lemma}\label{lemat}
If $R_k Q_l R_k=\gamma R_k$, $k \le l \le n-k$ holds for some $0<\gamma<1$ then $||Q_l-\gamma R_k||_{tr}=k\sqrt{(1-\gamma)(1+3\gamma)}+l-k$.
\end{lemma}
\ods
\pruf  Due to Lemma \ref{postac} we have
\beq
Q_l-\gamma R_k=\left[
\begin{array}{c|c|c}
\begin{array}{ccc}
  0 &  &    \\
   &  \ddots &  \\
   &  &
   0
\end{array}
 & \begin{array}{ccc}
  \sqrt{\gamma(1-\gamma)} &  &   \\
   &  \ddots &  \\
   &   & \sqrt{\gamma(1-\gamma)}
\end{array} & 0\\ \hline
 \begin{array}{ccc}
  \sqrt{\gamma(1-\gamma)} &  &  \\
   &  \ddots &  \\
    &  & \sqrt{\gamma(1-\gamma)}
\end{array} & \begin{array}{ccc}
  1-\gamma &  &    \\
   &   \ddots &  \\
   &    & 1-\gamma
\end{array}  & 0\\ \hline
0 & 0 & S
\end{array}\right].
\eeq
The matrix has a structure of a direct sum and we immediately obtain
\beqn
||Q_l-\gamma R_k ||_{tr}=k \Bigg|\Bigg|\left(
                                \begin{array}{cc}
                                  0 & \sqrt{\gamma(1-\gamma)} \\
                                 \sqrt{\gamma(1-\gamma)} & 1-\gamma \\
                                \end{array}
                              \right)
\Bigg|\Bigg|_{tr}+l-k=
k\sqrt{(1-\gamma)(1+3\gamma)}+l-k,\nonumber
\eeqn
where the last term is just the trace of $S$, which is of rank $l-k$.
\prufend
\ods
\begin{lemma}\label{pomocne}
Let the following hold for some $\gamma>0$ and the rank $l$ projection $Q_l$
\beq\label{prod}
(R_M\otimes R'_N) Q_l (R_M\otimes R'_N)=\gamma R_M\otimes R'_N.
\eeq
Then: \linia(i) \beq r(\mathrm{tr}_B Q_l)\ge M,\quad r(\tr_A Q_l)\ge N,\eeq
\linia
(ii)
\beqn
||\tr_B Q_l-\gamma N R_M||_{tr}\le MN \sqrt{(1-\gamma)(1+3\gamma)}+l-MN,\\
||\tr_A Q_l-\gamma M R'_N||_{tr}\le MN \sqrt{(1-\gamma)(1+3\gamma)}+l-MN \nonumber.
\eeqn
\end{lemma}
\pruf We prove the result for one of the partial traces, as the other case can be solved in an analogous way. (i) For the proof of the first part we write
for some arbitrary $\ket{\varphi}$ from $\calR_M$ (a subspace with projection $R_M$)
\beq
\bra{\varphi}\mathrm{tr}_B Q_l\ket{\varphi}=\sum_{\ket{\psi_i}\in \calR_N'}\bra{\varphi}\bra{\psi_i}Q_l \ket{\varphi}\ket{\psi_i}+\sum_{\ket{\psi_i}\in \calR '{}_N^{\perp}}\bra{\varphi}\bra{\psi_i}Q_l \ket{\varphi}\ket{\psi_i}.
\eeq
Recall now that Eq. (\ref{prod}) is equivalent to
\beq
\bra{\varphi_i}\bra{\psi_j}Q_l\ket{\varphi_s}\ket{\psi_m}=\delta_{is}\delta_{jm}\lambda
\eeq
with orthonormal basis $\ket{\varphi_i}$ and
$\ket{\psi_i}$ for $\calR_M$ and $\calR'_N$ respectively.
In virtue of this fact first sum is exactly $N\lambda$. Second sum is nonnegative since $Q_l$ is positive semi-definite. It thus certainly holds that
\beq \label{ineq}
\bra{\varphi}\mathrm{tr}_B Q_l\ket{\varphi}\ge N\lambda>0,\quad \forall_{\ket{\varphi}\in \calR_M}.
\eeq
Since matrix multiplication cannot increase rank, one has $r(\mathrm{tr}_B Q_l)\ge r(R_M \mathrm{tr}_B Q_l R_M)$. Thus it remains to show that Eq. (\ref{ineq}) implies $r(R_M \mathrm{tr}_B Q_l R_M)=r(R_M)$. Take $R_M=\sum_{i=1}^M \proj{\varphi_i}$, then  naturally $R_M\mathrm{tr}_B Q_l R_M=\sum_{i=1}^M \gamma_i \proj{\varphi_i}$, $\gamma_i \ge 0$. It is now sufficient to put this into Eq. (\ref{ineq}) to conclude that $\gamma_i>0$ for all $i$, which ends this part of  the proof.
(ii) The second assertion follows directly from Lemma \ref{lemat} since for any $G$, $H$ it holds that $||tr_B G-tr_B H||_{tr}\le ||G-H||_{tr}$ \cite{chuang-nielsen}.
\prufend
\ods\\
\textbf{Proof of Theorem \ref{konieczny}:} The result follows directly from Lemma \ref{pomocne} and the fact that for Hermitian matrices $A$, $B$ with eigenvalues $a_1\ge a_2 \ge \ldots \ge a_n$, $b_1\ge b_2 \ge \ldots \ge b_n$ respectively it holds that $||A-B||_{\mathrm{tr}}\ge\sum_{i=1}^n |a_i-b_i|$ \cite{Hayashi}.
\prufend\\
\ods

Theorem \ref{konieczny} can be also applied to $P_{d^2-l}=\jedynka_{d^2} - Q_l$, which provides the dual estimates.\ods\noindent
\begin{theorem}\label{dualne}
Let $P_{d^2-l}$ be a rank $d^2-l$ projection. Assume $(R_M\otimes R_N') P_{d^2-l} (R_M\otimes R_N') = (1- \gamma) R_M\otimes R_N'$ holds.
Let further $\tilde{x}_1\ge \tilde{x}_2\ge\dots$ be eigenvalues of $\mathrm{tr}_B P_{d^2-l}$, and $\tilde{y}_1\ge \tilde{y}_2\ge\dots$ of $\mathrm{tr}_A P_{d^2-l}$. Then
\beqn\label{ograniczenie2}
\sum_{i=1}^M |\tilde{x}_i-N(1-\gamma)|+\sum_{i=M+1}^{r\left(\mathrm{tr}_B P_{d^2-l}\right)} \tilde{x}_i\le MN\sqrt{\gamma(4-3\gamma)}+d^2-l-MN,\\
\sum_{i=1}^N |\tilde{y}_i-M(1-\gamma)|+\sum_{i=N+1}^{r\left(\mathrm{tr}_A P_{d^2-l}\right)} \tilde{x}_i\le MN\sqrt{\gamma(4-3\gamma)}+d^2-l-MN.\nonumber
\eeqn
\end{theorem}

Notice that Lemma \ref{postac} implies that eigenvectors of $Q$ must have a special form.\ods
  \begin{corollary}\label{wektory-postac}
 If $RQR=\gamma R$  holds then there exists the basis in which $k$ eigenvectors $\ket{\psi_i}$ of the projector $Q=\sum_{i=1}^q \proj{\psi_i}$ can be expressed as
 \beq\label{wektors}
 \ket{\psi_i}=\sqrt{\gamma}\ket{e_i}+\sqrt{1-\gamma}\ket{v_i},
 \eeq
 where $\inner{e_i}{e_j}=\delta_{ij}$, $\inner{v_i}{v_j}=\delta_{ij}$, and $\inner{e_i}{v_j}=0$.
 The code is then $R=\sum_i \proj{e_i}$ and its existence is equivalent to the existence of the  code $\hat{R}=\sum_i \proj{v_i}$ satisfying $\hat{R}Q\hat{R}=(1-\gamma) \hat{R}$. This  implies that for any projection $Q$ both $\gamma$ and $1-\gamma$ belong to $\Lambda_k(Q)$.
\end{corollary}\ods
Thus if one wants to find a code for $\gamma\ne 0,1$ one needs to find a basis for which Eq. (\ref{wektors}) holds.

Analysis of a concrete case in the next section will provide us with a proof of the following:
\ods
\begin{theorem}\label{odciete}
Existence of a product code for the noise model $U=P-Q$ for $\lambda=1-2\gamma$ does not necessarily imply the existence of a product code for $\lambda=2\gamma-1$. In other words, there are cases when $\gamma\in\Lambda_{m\otimes n}$ but $1-\gamma\notin\Lambda_{m\otimes n}$.
\end{theorem}
\ods
This result provides sharp distinction between the standard and the product numerical range.

In Ref. \cite{demianowicz-2012} it was shown that in the case of qutrit inputs ($d=3$) a zero entropy code is unique in a sense that there are no codes for the noise model Eq. (\ref{postac-u}) simultaneously for $\lambda=+1$ and $\lambda=-1$. Higher dimensional codes were not considered there. With the  above  results in hands we can prove that this uniqness is stronger. Namely, we have\ods
\begin{observation}\label{only}
If for $d=3$ there exists a product $2\otimes 2$ decoherence free subspace then there exists {\it no} higher entropy code for this system.
\end{observation}\ods
\pruf We will prove the result for a four dimensional projection operator $Q_4$ since the result is proved in a similar manner for a five dimensional projection. We only show the part which was not proved in Ref. \cite{demianowicz-2012}. Suppose the equality $(S\otimes S' )Q_4 (S\otimes S')=S\otimes S'$ holds. This means that $Q_4=S\otimes S'$. The following equality $(R\otimes R' )Q_4 (R\otimes R')=\gamma R\otimes R'$ with $0<\gamma<1$ cannot then hold since this requires $RSR=\tilde{\gamma}R$ and $R'S'R'=\tilde{\gamma}'R'$ with $\tilde{\gamma}\tilde{\gamma}'=\gamma$.  It is impossible since this implies that either $\gamma$ or $\gamma'$ (or both) is not equal to one, which stays in contradiction to Lemma \ref{postac} (applying the theorem to $S$ or $S'$ we see that they must be at least four dimensional, which contradicts the assumption $d=3$).

 Suppose now that $(R\otimes R') Q_4 (R\otimes R')=0$ holds. This implies that $P_5:=\jedynka_9-Q_4=R\otimes R'+\proj{\xi}$ with $\xi \perp R\otimes R'$. If there existed higher entropy code, it would hold that $(T\otimes T') (R\otimes R'+\proj{\xi})(T\otimes T')=\gamma T\otimes T'$ for some two dimensional projections $T$, $T'$. Hence $(T\otimes T') \proj{\xi} (T\otimes T')$ must be a nonzero vector if there is to be nonzero $\gamma$. Otherwise the operators $TRT$ and $T'R'T'$ would have to be proportional to $T$ and $T'$ respectively which is impossible.  This is a two qubit problem because of two dimensional projections so we rewrite it for clarity as $E\otimes F+\proj{\Psi}=\gamma \jedynka\otimes\jedynka$. Thus $G:=\gamma \jedynka\otimes\jedynka-E\otimes F$ must be one dimensional. Let $(e_1,e_2)$ and $(f_1,f_2)$ be the eigenvalues of $E$ and $F$ respectively. The eigenvalues of $G$ then read $(\gamma-e_1f_1,\gamma-e_1f_2,\gamma-e_2f_1,\gamma-e_2f_2)$. Exactly three of these eigenvalues have to be equal to zero, which is impossible.
\prufend
\ods
\\
\indent
Finally, notice that the whole reasoning from this section can be applied to an arbitrary noise model of the form $U=P+ \eksp^{\uroj\beta}Q$.

\section{Applications}
In what follows, we will be mainly interested in finding bipartite product higher rank numerical ranges of operators acting on $\mathbb{C}^d\otimes \mathbb{C}^d$. One example will be devoted to the case of different local dimensions.
\subsection{Random swap}
Among all unitary matrices of the form (\ref{postac-u}), $SWAP$ is the quantum gate probably most often used in the theory of quantum information.
 It can be written as $SWAP_{d\leftrightarrow d}=P_{sym.}-P_{asym.}:=V_d$, where $P_{sym.}$, $P_{asym.}$ are the projections onto symmetric and antisymmetric subspaces respectively. Its action on pure states is given by $V_d \ket{\phi}\ket{\psi}=\ket{\psi}\ket{\phi}$.

This kind of noise can be approached directly without resorting to the results presented above. We need to solve
\beq\label{swapa}
R\otimes R' V_d R\otimes R'=\lambda R\otimes R'
\eeq
with $R=\sum_{i=1}^M \proj{\varphi_i}$ and $R'=\sum_{j=1}^N \proj{\psi_j}$. It is equivalent to
\beq
\bra{\varphi_i}\bra{\psi_j} V_d \ket{\varphi_k}\ket{\psi_l}=\lambda \delta_{ik}\delta{jl},
\eeq
which in this case results in
\beq
\inner{\varphi_i}{\psi_l}\inner{\varphi_k}{\psi_j}^*=\lambda \delta_{ik}\delta_{jl}.
\eeq
As one can realize from the above, $\lambda=0$ is the only possible value.
The corresponding eigenvectors of $R$ and $R'$ must obey $\varphi_i \perp \psi_j$, thus
\beq
\kan_{2\otimes 2}(V_3)=\emptyset
\eeq
and
\beq
\Lambda_{M\otimes N}(V_d)=\{0\},\quad M+N\le d, \quad d\ge 4.
\eeq
In the latter case with $M=N=2$ we could take for example $R=\proj{0}+\proj{1}$ and $R'=\proj{2}+\proj{3}$.

The above also means that we have:
\beq
\kan_{M\otimes M}^{\mathrm{symm.}}(V_d)=\emptyset.
\eeq

Notice that the local numerical range (see Section \ref{prod-higher}) provides a quick rough upper bound $\langle 0,1\rangle$ on the $M\otimes N$ range. This is an example of an application of the Property \ref{zawieranie-w-mniejszych}.
Another bound narrowing previously mentioned one can be readily obtained from Fact \ref{prod-w-C}, which in this case does not involve optimization. Taking the trace of both sides of Eq. (\ref{swapa}) and exploiting the fact $\tr V_d (A\otimes B)=\tr AB$ we obtain
$\tr RR'=MN\lambda$.
Utilizing now the property (following from the H\"older's inequality \cite{lutkepohl-handbook}) $\tr AB \le \tr A$ holding for $A\ge 0$ and $\mathbb{I}\ge B \ge 0$ and the fact that $\tr AB \ge 0$ for positive semi--definite $A$ and $B$, we obtain
$
0\le MN \lambda \le \min \left\{ M,N \right\},
$
which finally leads to
\beq
0\le \lambda\le \min\displaystyle \left\{\frac{1}{M},\frac{1}{N}\right\},
\eeq
giving a significant improvement over the previous bound.

Presented reasoning shows that in some cases direct exploitation of properties of $U$ may prove very useful.
\subsection{Other kinds of noise}
\subsubsection{The $3\otimes 3$ systems}
We start with $d=3$ examples.\\
{\it Example 1.}\\
 We will establish an outer bound for  $\Lambda_{2\otimes 2}$ of a projector $Q(\alpha)$
having the following eigenvectors
\beqn
\ket{\psi_1}= (\sqrt{\alpha}\ket{0}+\sqrt{1-\alpha}\ket{2})\otimes\ket{0}   ,\\\
\ket{\psi_2}= \ket{0} \otimes (\sqrt{\alpha}\ket{1}+\sqrt{1-\alpha}\ket{2}) ,\\\
\ket{\psi_3}= \ket{1}  \otimes(\sqrt{\alpha}\ket{0}+\sqrt{1-\alpha}\ket{2}),\\\
\ket{\psi_4}= (\sqrt{\alpha}\ket{1}+\sqrt{1-\alpha}\ket{2})\otimes\ket{1}
\eeqn
with $0<\alpha<1$.
From Corollary \ref{wektory-postac} we see that $\alpha\in \Lambda_{2\otimes 2}(Q(\alpha))$ as all above vectors are of the required form $\sqrt{\alpha}\ket{e_i}+\sqrt{1-\alpha}\ket{v_i}$. Notice that it also follows that $\alpha \in \kan_{2\otimes 2}^{\mathrm{symm.}}(Q(\alpha))$.

\noindent Simple calculation yields
\beq
X\equiv \tr_A\; Q(\alpha)=\tr_B\; Q(\alpha)=
\left(
  \begin{array}{ccc}
   1+\alpha & 0 & \sqrt{\alpha(1-\alpha)} \\
   0&  1+\alpha & \sqrt{\alpha(1-\alpha)} \\
    \sqrt{\alpha(1-\alpha)} &\sqrt{\alpha(1-\alpha)} &  2-2\alpha\\
  \end{array}
\right)
\eeq
and
$Y\equiv \tr_A\; (\jedynka_9-Q(\alpha))=\tr_B\; (\jedynka_9-Q(\alpha))=3\jedynka_3-X$.
The eigenvalues are: $x_1=2$, $x_2=1+\alpha$, $x_3=1-\alpha$ and $\tilde{x}_1=2+\alpha$, $\tilde{x}_2=2-\alpha$, and $\tilde{x}_3=1$. Setting $M=N=2$ we get from Theorems \ref{konieczny} and \ref{dualne} two inequalities on $\gamma$:
\beqn
|2-2\gamma|+|1+\alpha-2\gamma|+1-\alpha\le 4\sqrt{(1-\gamma)(1+3\gamma)},\\
|2+\alpha-2(1-\gamma)|+|2-\alpha-2(1-\gamma)|+1\le 4\sqrt{\gamma(4-3\gamma)}+1,
\eeqn
from which we get
\beq\label{przedzial}
\gamma \in \Lambda_{2\otimes 2}\left(Q(\alpha)\right)\subseteq\displaystyle\Bigg\langle \frac{2}{3}\left(1-\sqrt{1-\frac{3}{16}\alpha^2}\right),\frac{1}{3}\left(1+\sqrt{4-3\left(\frac{1-\alpha}{2}\right)^2}\right)\Bigg\rangle,
\eeq
For each value of $\alpha$ we thus get an interval outerbounding $\Lambda_{2\otimes 2}$.
One realizes that this example already provides the evidence in favor of Theorem \ref{odciete}.

\begin{figure} [tbp]
\centerline{\epsfig{file=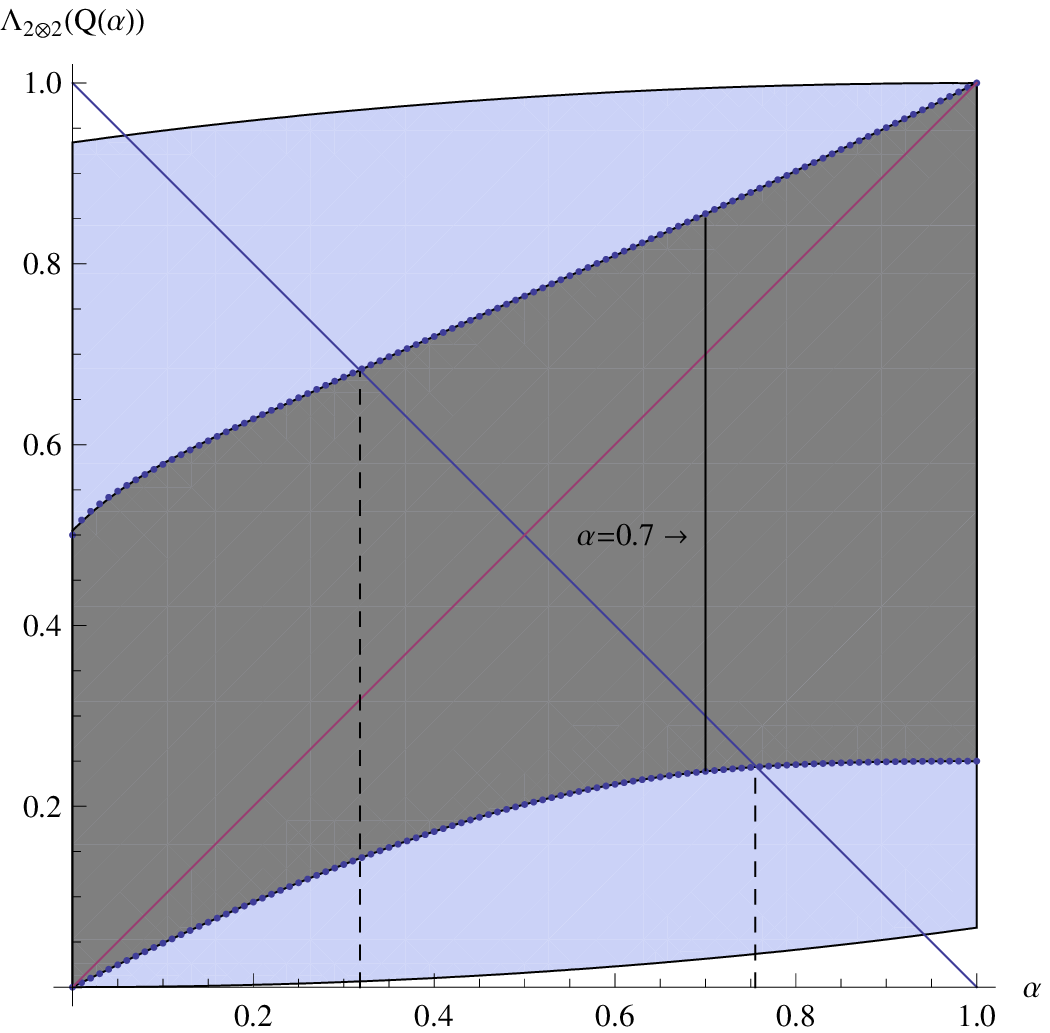, width=8.2cm,height=8cm}} 
\vspace*{13pt}
\fcaption{\label{oporny-przypadek-rys} The light shaded region represents the bound given by Eq. (\ref{przedzial}) in the whole range of $\alpha$, the dark grey region is the bound given by Eq. (\ref{optymalizowany}) (the dots are for $\lambda_{\uparrow,\downarrow}(\alpha)$). We have also put the section bounding the product range for an exemplary value $\alpha=0.7$. Additionally, we plotted the lines $\alpha$ and $1-\alpha$. While the former value always belongs to the product range (as we have discussed in the main text) there is a range of $\alpha$, where the latter lies outside  it with certainty as it is cut off by the bound, this cut off is represented by the dashed lines.}
\end{figure}

Using Observation \ref{prod-w-C}  this bound can be improved extending the region where $\alpha$ belongs to the range but $1-\alpha$ does not. As we have noted in the previous section, this usually requires optimization (in opposite to the application of Theorems \ref{konieczny} i \ref{dualne} as above). We need to calculate
\beq
\lambda_{\uparrow}(\alpha):= \frac{1}{4}\min _{U\otimes V}  \tr [(U\otimes V) (P_0\otimes P_0) (U^{\dagger}\otimes V^{\dagger})Q(\alpha)]\eeq
and
\beq
\lambda_{\downarrow}(\alpha):=\frac{1}{4}\max _{U\otimes V} \tr [(U\otimes V) (P_0\otimes P_0) (U^{\dagger}\otimes V^{\dagger})Q(\alpha)], \eeq
where $P_0=\proj{0}+\proj{1}$,
which give rise, according to Observation \ref{prod-w-C}, to the bound
\beq \label{optymalizowany}
\Lambda_{2\otimes 2}(Q(\alpha))\subseteq [\lambda_{\uparrow}(\alpha);\lambda_{\downarrow}(\alpha)].
\eeq
With this aim we parameterized  $U$ and $V$ according to Ref. \cite{kus-zyczkowski} and then, for different values of $\alpha \in (0,1)$ with the step $\delta \alpha =0.01$, computed $\lambda_{\uparrow}$ and $\lambda_{\downarrow}$. In this case the same bound is valid for $\kan_{2\otimes 2} ^{\mathrm{symm.}}$.  The results of this part of the paper are shown in Fig.  \ref{oporny-przypadek-rys} (see also Section \ref{dyskusja}).

\noindent{\it Example 2.}\\
Consider now the following operator \cite{PH-ML2000}
\beq
Q_l=\sum_{i<j} \proj{\psi_{ij}}+P_d^{+},
\eeq
where
\beq
\ket{\psi_{ij}}=\frac{1}{\sqrt{d}}\left(\sqrt{a}\ket{ij}+\sqrt{a^{-1}}\ket{ji}\right),\quad a+a^{-1}=d.
\eeq
We set $d=3$. Then,  we obtain an outer bound
\beq
\Lambda_{2\otimes 2}\subseteq\left[\frac{4-\sqrt{13}}{6};\frac{3+\sqrt{33}}{9}\right]\approx [0.0657;0.9716].
\eeq
At the same time, application of Fact \ref{prod-w-C} gives us better estimates:
\beq
\Lambda_{2\otimes 2}\subseteq [ 0.1788;0.7378].
\eeq

We have also used Observation \ref{prod-w-C} to estimate bounds on symmetric product range and we have obtained that
\beq
\Lambda_{2\otimes 2}^{\mathrm{symm.}}\subseteq [ 0.4166;0.5556].
\eeq

We were not able to verify whether any of the values inside these intervals actually belong to the (symmetric) product higher rank numerical range of $Q$.
\subsubsection{The $4\otimes 4$ systems}
Now we give an example of the projection $Q$ for which the numerical range $\Lambda_{2\otimes 2}$ can be found. New analytical technique will be introduced with this aim. Let the eigenvectors of $Q(\alpha)$ be
\beq\label{wektory-wlasne1}
\ket{\Psi_1}=\sqrt{\alpha}\ket{00}+\sqrt{1-\alpha}\ket{22},
\eeq
\beq
\ket{\Psi_2}=\sqrt{\alpha}\ket{01}+\sqrt{1-\alpha}\ket{23},
\eeq
\beq
\ket{\Psi_3}=\sqrt{\alpha}\ket{10}+\sqrt{1-\alpha}\ket{32},
\eeq
\beq\label{wektory-wlasne2}
\ket{\Psi_4}=\sqrt{\alpha}\ket{11}+\sqrt{1-\alpha}\ket{33}.
\eeq
Theorem \ref{postac} leads to the conclusion that both $\alpha$ and $1-\alpha$ belong to $\Lambda_{2\otimes 2}(Q(\alpha))$. Assume $1-\alpha\ge\alpha$. Consider now projectors $R$ and $R'$ spanned respectively by
\beqn\label{wybor1}
\ket{\tilde{\phi}_1}=\sqrt{1-\beta}\ket{0}+\sqrt{\beta}\ket{2},\quad
\ket{\tilde{\phi}_2}=\sqrt{1-\beta}\ket{1}+\sqrt{\beta}\ket{3}
\eeqn
and
\beqn\label{wybor2}
\ket{\tilde{\psi}_1}=\ket{2},\quad
\ket{\tilde{\psi}_2}=\ket{3}
\eeqn
with $\beta\in\langle 0,1 \rangle$.
One can easily check that with such choice of projectors we can get $\gamma=\beta(1-\alpha)$ and thus varying $\beta$ over the whole range
 in turn that $\gamma\in\langle0,1-\alpha\rangle$. We will now show that this is best what one can do and this interval represents $\Lambda_{2\otimes 2}(Q(\alpha))$.

With this aim recall once again that if $\gamma\in\Lambda_{2\otimes 2}$ then there exist projectors $R=\proj{\phi_1}+\proj{\phi_2}$ and $R'=\proj{\psi_1}+\proj{\psi_2}$ such that
$\bra{\phi_i}\bra{\psi_j}Q_4\ket{\phi_k}\ket{\psi_l}=\gamma \delta_{ik}\delta_{jl}.$
Naturally $\langle\phi_i|\phi_j\rangle=\delta_{ij}$ and $\langle\psi_i|\psi_j\rangle=\delta_{ij}$.
Let the vectors be decomposed as follows
\beqn\label{wektory}
\ket{\phi_1}=\ket{e,e'},\quad
\ket{\phi_2}=\ket{f,f'},\quad e,e',f,f'\in\mathbb{C}^2,\\
\ket{\psi_1}=\ket{E,E'},\quad
\ket{\psi_2}=\ket{F,F'},\quad E,E',F,F'\in\mathbb{C}^2,
\eeqn
where $\ket{i,j}$ is understood as the direct sum $\ket{i}\oplus\ket{j}$, \tzn, we use the isomorphism $\mathbb{C}^4\cong \mathbb{C}^2\oplus \mathbb{C}^2$.
Let now $p,p',q,q',P,P',Q,Q'$ be variables in $\mathbb{C}^2$, which can take values as follows $p,q=e,f$; $p',q'=e',f'$; $P,Q=E,F$; $P',Q'=E',F'$ in combinations consistent with Eq. (\ref{wektory}).
After a direct calculation we arrive at
\beq\label{transformed}
\bra{p,p'}\bra{P,P'}Q_4\ket{q,q'}\ket{Q,Q'}=\left(\sqrt{\alpha}\bra{p}\bra{P}+\sqrt{1-\alpha}\bra{p'}\bra{P'}\right)
\left(  \sqrt{\alpha}\ket{q}\ket{Q}+\sqrt{1-\alpha}\ket{q'}\ket{Q'}  \right),
\eeq
which must be zero whenever $(p,p')\ne(q,q')$ or $(P,P')\ne(Q,Q')$ and $\gamma$ otherwise.
Recalling what values can be taken by the variables and orthogonality conditions we thus conclude
\begin{lemma}
The number $\gamma$ belongs to $\Lambda_{2\otimes 2}(Q(\alpha))$ if and only if there exists a set of four vectors from $\mathbb{C}^2\otimes\mathbb{C}^2$
\beqn
\ket{\chi_1}=\sqrt{\alpha}\ket{e}\ket{E}+\sqrt{1-\alpha}\ket{e'}\ket{E'},\\
\ket{\chi_2}=\sqrt{\alpha}\ket{e}\ket{F}+\sqrt{1-\alpha}\ket{e'}\ket{F'},\\
\ket{\chi_3}=\sqrt{\alpha}\ket{f}\ket{E}+\sqrt{1-\alpha}\ket{f'}\ket{E'},\\
\ket{\chi_4}=\sqrt{\alpha}\ket{f}\ket{F}+\sqrt{1-\alpha}\ket{f'}\ket{F'}\\
\eeqn
such that
\beqn
||\ket{i}||^2+||\ket{i'}||^2=1,\quad i=e,f,E,F,\\
\langle e|f   \rangle+\langle e'|f'   \rangle=0,\quad
\langle E|F   \rangle+\langle E'|F'   \rangle=0
\eeqn
and
\beqn
\langle \chi_i|\chi_j   \rangle=\gamma\delta_{ij}.
\eeqn
\end{lemma}
This observation is crucial as now we can bound $\gamma$. Writing
$||\ket{i}||$ as $||i||$ for clarity, we have for example for
$\ket{\chi_1}$ \beqn &\sqrt{\gamma}= \sqrt{\langle \chi_1|\chi_1
\rangle}= ||\chi_1||&= ||\sqrt{\alpha}\ket{e}\ket{E}+
\sqrt{1-\alpha}\ket{e'}\ket{E'} ||\non
&&\le\sqrt{\alpha}\;||e||\cdot ||E ||+\sqrt{1-\alpha}\;||e'||\cdot
||E'|| \non &&\le \sqrt{1-\alpha}\;\left(||e||\cdot ||E||    +
||e'||\cdot ||E'|| \right)\non
&&=\sqrt{1-\alpha}\;\left[||e||,||e'||\right]\cdot
\left[||E||,||E'||\right] \non &&\le\sqrt{1-\alpha}\;
 \Big|\Big|\left[||e||,||e'||\right]
 \Big|\Big|\cdot \Big|\Big|\left[||E||,||E'||\right]
 \Big|\Big|\non
 &&=\sqrt{1-\alpha}\;\sqrt{||e||^2+||e'||^2}
 \sqrt{||E||^2+||E'||^2}\non
 &&=\sqrt{1-\alpha},\nonumber
\eeqn which finally leads to \beq \Lambda_{2\otimes
2}(Q(\alpha))=\langle 0, 1-\alpha \rangle. \eeq Let us justify all
the important steps in the estimation for $\gamma$: (i) the first
inequality follows from the triangle inequality, (ii) the second
inequality uses $\alpha\le 1-\alpha$, (iii) the third is the
Cauchy-Schwarz inequality, (iv) the last equality makes use of
normalization.

Proceeding in a similar manner in the case of $\alpha \ge 1-\alpha$ we find that in this region the product numerical range is $\Lambda_{2\otimes 2}(Q(\alpha))=\langle 0, \alpha \rangle$.

Interestingly, inclusion of sets from Observation \ref{prod-w-C} becomes an equality in this case so the same region can be found by numerical optimization of the bounds on the local $R\otimes R'$--numerical range of $Q(\alpha)$ (see previous examples).

\begin{figure} [htbp]
\centerline{\epsfig{file=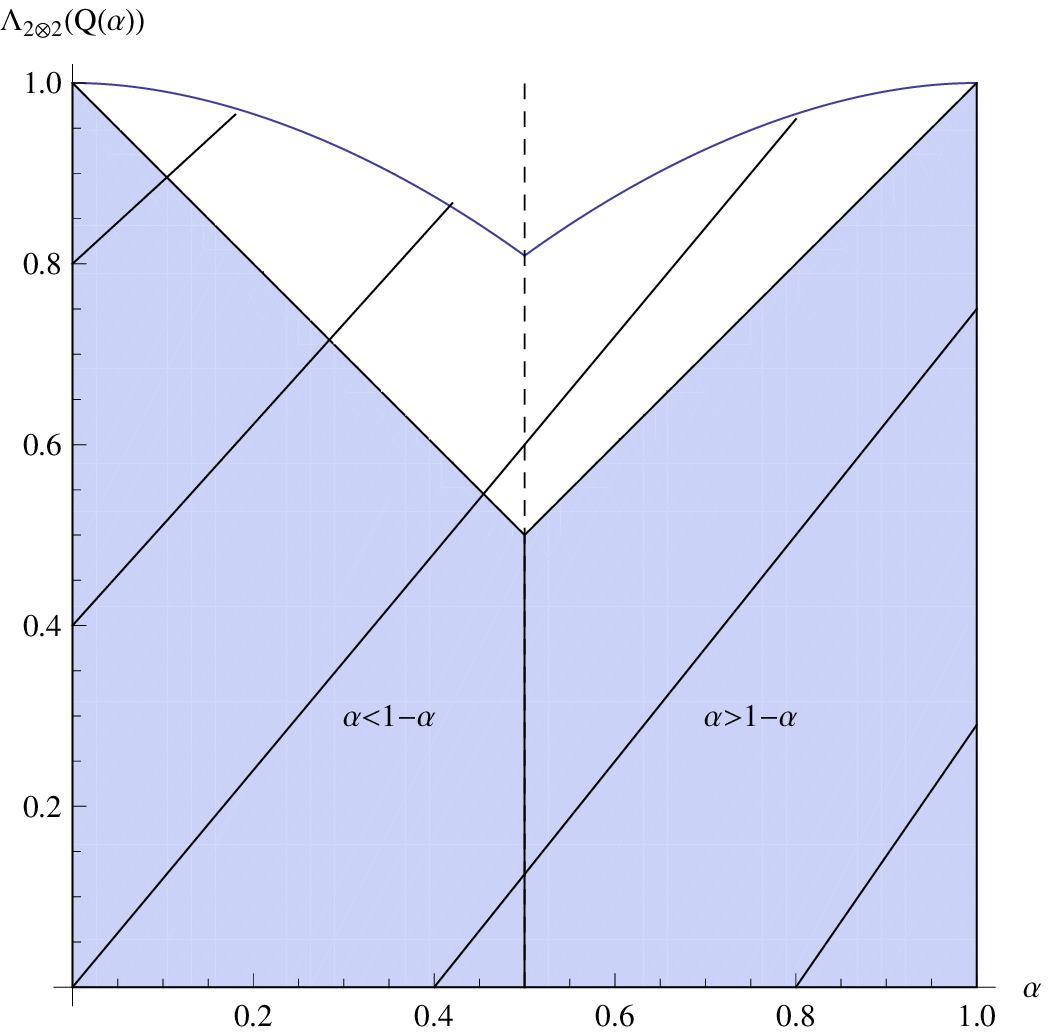, width=8.2cm}} 
\vspace*{13pt}
\fcaption{\label{2x4} The shaded region is the $2\otimes 2$ range of $Q(\alpha)$ given by Eqs (\ref{wektory-wlasne1}-\ref{wektory-wlasne2}). The hatched region is the bound from Theorems \ref{konieczny} and \ref{dualne}. }
\end{figure}

It also turns out that in this case
\beq \kan_{2\otimes 2}^{\mathrm{symm.}} (Q(\alpha))=\kan_{2\otimes 2} (Q(\alpha)),\eeq
which can be shown to hold with vectors
\beqn
\ket{\phi_1}=\ket{\psi_1}=\sqrt{1-\beta}\ket{0}+\uroj \sqrt{\beta}\ket{2},\quad
\ket{\phi_2}=\ket{\psi_2}=\sqrt{1-\beta}\ket{1}+\uroj \sqrt{\beta}\ket{3}.
\eeqn
\indent The results of this section are summarized in Fig. \ref{2x4}, in which we have also plotted the bound stemming from Theorems \ref{konieczny} and \ref{dualne}.

\subsubsection{The $2\otimes 4$ systems}
Consider the following projector acting on $\mathbb{C}^2\otimes\mathbb{C}^4$
\beq\label{kubit-kuqart}
Q(\gamma)=\proj{0}\otimes Q_1+\proj{1}\otimes Q_2(\gamma)
\eeq
with
\beq
Q_1=\proj{0}+\proj{1},
\quad
Q_2(\gamma)=\proj{\eta_1}+\proj{\eta_2},
\eeq
\beqn
\ket{\eta_1}=\sqrt{\gamma}\ket{0}+\sqrt{1-\gamma}\ket{2},\quad \ket{\eta_2}=\sqrt{\gamma}\ket{1}+\sqrt{1-\gamma}\ket{3}.
\eeqn
From Property \ref{wspolny-projektorow} it follows that we must find $\Lambda_2^{\mathrm{comm.}}(Q_1,Q_2(\gamma))$. Taking $\alpha=1/2$ in Property \ref{ograniczenia-na-common}, we obtain $\Lambda_2^{\mathrm{comm.}}(Q_1,Q_2(\gamma))\subseteq \kan_2(Z(\gamma))$ with
\beq Z(\gamma)=\frac{1}{2}
 \left(
   \begin{array}{cccc}
     1+\gamma & 0 & \sqrt{\gamma(1-\gamma)} & 0 \\
     0 & 1+\gamma & 0 & \sqrt{\gamma(1-\gamma)} \\
     \sqrt{\gamma(1-\gamma)} &0 & 1-\gamma & 0 \\
     0 & \sqrt{\gamma(1-\gamma)} & 0 & 1-\gamma \\
   \end{array}
 \right).
 \eeq
 By inspection, we find the eigenvalues of $Z(\gamma)$ to be $\frac{1}{2}(1-\sqrt{\gamma}),\frac{1}{2}(1-\sqrt{\gamma}),
\frac{1}{2}(1+\sqrt{\gamma}),\frac{1}{2}(1+\sqrt{\gamma})$.
From the theory of higher rank numerical range (see Section \ref{higher-approach}) it immediately follows that
\begin{figure} [htbp]
\centerline{\epsfig{file=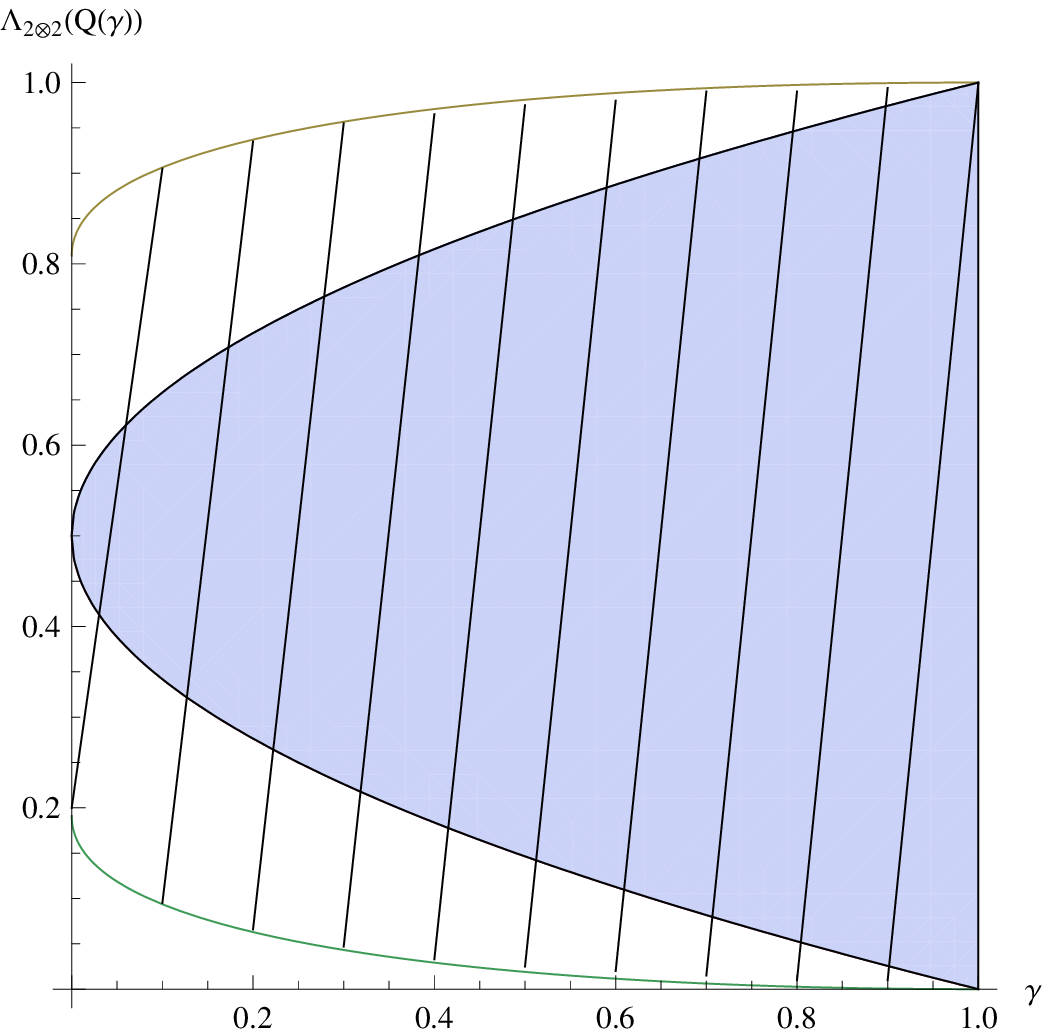, width=8.2cm}} 
\vspace*{13pt}
\fcaption{The shaded region is the product higher rank numerical range of $Q(\gamma)$ given by Eq. (\ref{kubit-kuqart}). The hatched region in which it is included is the bound stemming from the application of Theorems \ref{konieczny} and \ref{dualne}.}\label{z-boundem}
\end{figure}
\beq\label{rank-z}
\Lambda_2(Z(\gamma))=\left[\frac{1}{2}(1-\sqrt{\gamma}),
\frac{1}{2}(1+\sqrt{\gamma})\right].
\eeq
It remains to show that the above represents the sought common numerical range (and in consequence the product numerical range), that is all $\lambda\in \kan_2(Z(\gamma))$ are achievable in the sense of fulfilling the following equations: (i) $RQ_1R=\lambda R$ and (ii) $RQ_2(\gamma)R=\lambda R$ for some $R$. Indeed, take $R$ to be projecting on the subspace spanned by the following two vectors:
\beqn
\ket{\xi_1}=\sqrt{\lambda}\ket{0}+\eksp^{\uroj \beta}\sqrt{1-\lambda}\ket{2},\quad
\ket{\xi_2}=\sqrt{\lambda}\ket{1}+\eksp^{\uroj \beta} \sqrt{1-\lambda}\ket{3}.
\eeqn
Then, by varying $\beta$ in the range $[0,\pi]$, we can get any $\lambda$ from the desired interval.

\begin{figure} [htbp]
\centerline{\epsfig{file=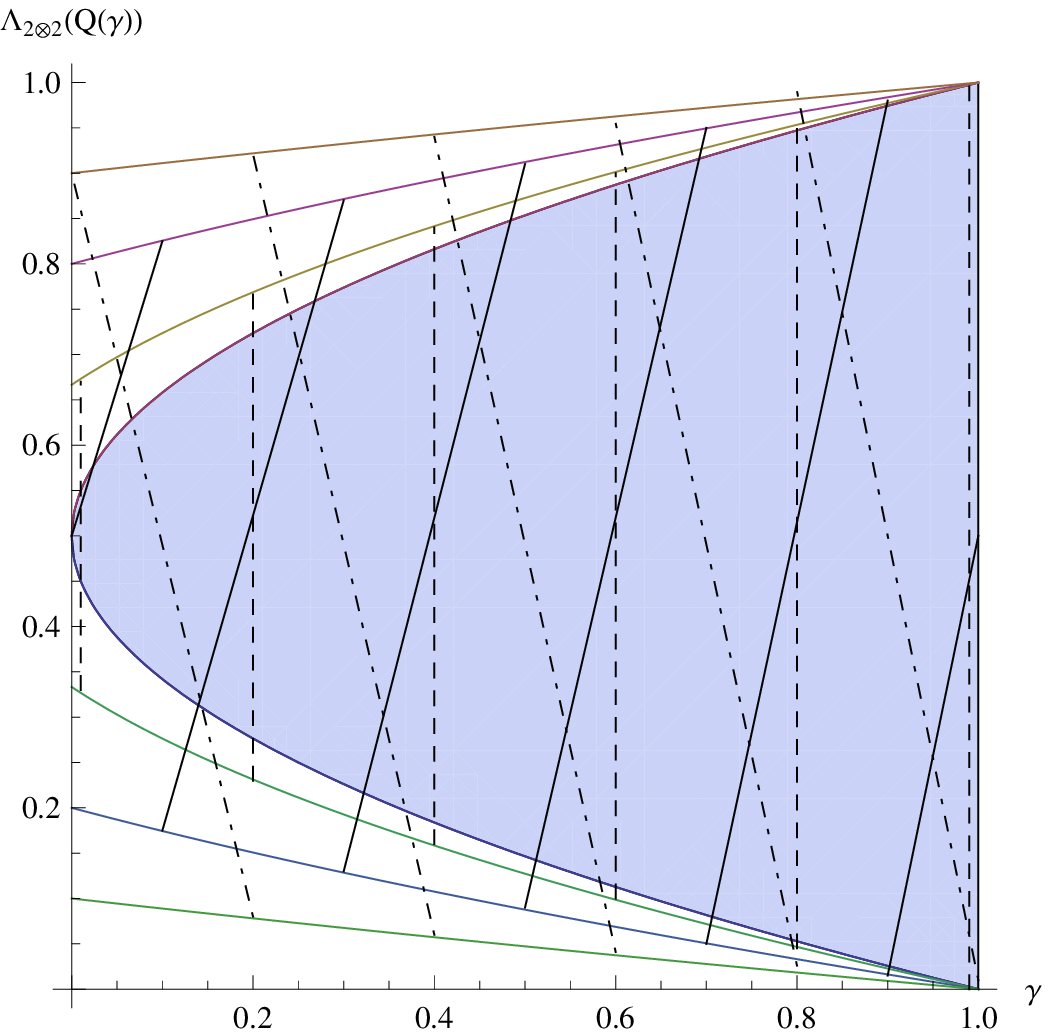, width=8.2cm}} 
\vspace*{13pt}
\fcaption{The inner shaded region corresponds to $\alpha=1/2$ in Property \ref{ograniczenia-na-common} and represents the true $2\otimes 2$ rank of $Q(\gamma)$ given by Eq. (\ref{kubit-kuqart}). The region with dashed lines corresponds to $\alpha=1/3$, solid lines --- $\alpha=1/5$, dot dashed lines --- $\alpha=1/10$.  In the limit $\alpha\to 0$ we will obviously get $[0,1]$.}\label{z-boundami}
\end{figure}

We have plotted the range in Fig. \ref{z-boundem} along with the bound stemming from Theorems \ref{konieczny} and \ref{dualne}. In Fig. \ref{z-boundami} we show application of Property \ref{ograniczenia-na-common} for values $\alpha=1/2,1/3,1/4,1/5,1/10$.

We have also verified by optimization that Observation \ref{prod-w-C} gives exactly (\ref{rank-z}) as the outer bound.
\section{Reverse problem --- a toy model}\label{toy}
We conclude with an example of a different type of noise. Unitary $U$ which we will consider have a regularly distributed spectrum and can hardly be called generic, nevertheless, as they can be treated analytically, we believe that the examples may be an important toy model for future work on the product numerical range. We concentrate on  highest entropy codes, which correspond to $\lambda=0$, but the observation can be generalized to other compression values. Our examples may be considered as an illustration of the inverse problem to the one considered so far. That is
\linia {\bf Problem.}
{\it  Given a product code $R\otimes R'$ construct a class of unitaries $U$ for which}
\beq
(R\otimes R') U (R\otimes R')=0.
\eeq
We will focus on $2\otimes 2$ codes and consider $d=4$ case. Using the terminology introduced in Section \ref{prod-higher} this problem can be rephrased as follows: {\it given an element of a product codes set of some unitary operator $U$ corresponding to the compression value $\lambda=0$, find an exemplary form of $U$}. Not surprisingly homogenous equation on $U$ is very different from homogenous equations on projectors considered in \cite{demianowicz-2012} so we need a different approach.

In our case we have $P=R\otimes R'$, \tzn,\; the sum $\sum_{i\in J} \proj{\psi_i}$ is a product projector ($J$ is a set of indices). It is worth stressing that $\ket{\psi_i}$ do not have to be product itself --- they only need to span four dimensional product subspace. We now need to construct spectrum of $U=\sum_i z_i\proj{v_i}$ so that $\lambda=0$ is a compression value. This can be easily done if we recall the well known identity
\beq\label{identity}
1+\omega+\omega^2=0;\quad \omega=\eksp^{ \frac{2\pi}{3}\uroj}.
\eeq
We choose
\beqn\label{w-wlasne}
&&\mathrm{spec}(U)=\{z_i\}_{i=1}^{16}=\nonumber\\&&\{ \eksp^{ \uroj\xi_1},\omega\eksp^{\uroj \xi_1},\omega^2\eksp^{ \uroj\xi_1},\eksp^{ \uroj\xi_2},\omega\eksp^{\uroj \xi_2},\omega^2\eksp^{ \uroj\xi_2},\eksp^{ \uroj\xi_3},\omega\eksp^{\uroj \xi_3},\omega^2\eksp^{ \uroj\xi_3},\eksp^{ \uroj\xi_4},\omega\eksp^{\uroj \xi_4},\omega^2\eksp^{ \uroj\xi_4}, \eksp^{ \uroj\xi_5},\eksp^{ \uroj\xi_6},\eksp^{ \uroj\xi_7},\eksp^{ \uroj\xi_8}       \}\nonumber
\eeqn
with arbitrary $\xi_i\ne \xi_j$ (at least for $i,j=1,2,3,4$). As it was mentioned earlier in Section \ref{higher-approach}, we can construct triangles from the eigenvalues, that is we can now
take $\delta_m=\{ \eksp^{ \uroj\xi_m},\omega\eksp^{\uroj \xi_m},\omega^2\eksp^{ \uroj\xi_m}   \}$ and due to Eq. (\ref{identity}) this means that indeed $\lambda=0$ is a compression value since we can set the numbers $\alpha$ to be all equal to $1/3$ (see Eq. (\ref{compress})). It remains now to properly choose eigenvectors. This construction uses the same arithmetic identity. We take $\ket{v_i}$ to be such that all terms in superposition besides the first one cancel due to Eq. (\ref{identity}) when added with coefficients $1/\sqrt{3}$ stemming from Eq. (\ref{weks}). The states remaining after the summation should sum to a product projector. We have explicitly in a closed form
\beq
\left(
  \begin{array}{c}
    \ket{v_i} \\
    \ket{v_{i+1}} \\
    \ket{v_{i+2}} \\
  \end{array}
\right):=\frac{1}{\sqrt{3}}
\left(
  \begin{array}{ccc}
    1 & 1 & 1 \\
    1 & \omega & \omega^2 \\
    1 & \omega^2 & \omega^4 \\
  \end{array}
\right)
\left(
  \begin{array}{c}
    \ket{u_i} \\
    \ket{u_{i+1}} \\
    \ket{u_{i+2}} \\
  \end{array}
\right),\quad i=1,4,7,10,
\eeq
with orthonormal states $\ket{u_i}$ such that
\beq\label{sumowanie}
\sum_{i=1,4,7,10} \proj{u_i}=R\otimes R'.
\eeq
We could, for example, take two--qubit Bell states embedded in a two ququart space, \tzn,
\beqn\label{eigen-choice}
\ket{u_{1,4}}=\frac{1}{\sqrt{2}}\left( \ket{00}\pm\ket{11}  \right), \quad
\ket{u_{7,10}}=\frac{1}{\sqrt{2}}\left( \ket{01}\pm\ket{10}  \right).
\eeqn
The rest of eigenvectors may be chosen arbitrarily with the only restriction that all $\ket{v_i}$ are orthonormal.

Let us summarize for clarity all the elements.
We have
\beqn\label{sum1}
\sum_{i=0}^2 \frac{1}{3}z_{i+k}=0,\quad k=1,4,7,10,
\eeqn
which ensures that $\lambda=0$ belongs to numerical range.
\begin{figure} [htbp]
\centerline{\epsfig{file=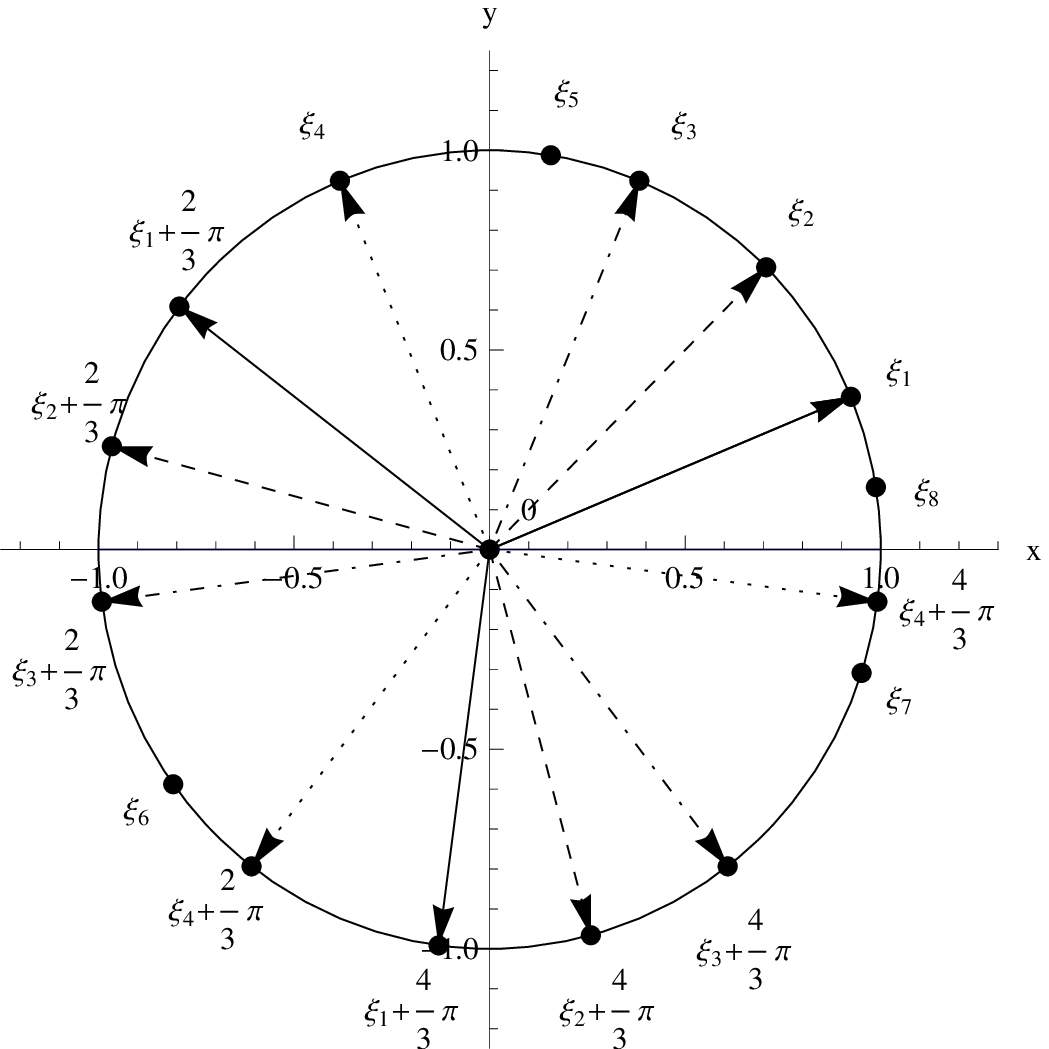, width=8.2cm}} 
\vspace*{13pt}
\fcaption{\label{zegarek1} Distribution of eigenvalues from Eq. (\ref{w-wlasne}) (for clarity we have written only phases). The arrows represent the summations of Eq. (\ref{sum1}). }
\end{figure}
We then assume that Eq. (\ref{sumowanie}) holds so we can define
\beqn
\ket{\psi_k}:=\frac{1}{\sqrt{3}}\sum_{m=0}^2\ket{v_{m+k}}=\ket{u_k},\quad k=1,4,7,10,
\eeqn
which properly sums to a product code.

We also propose an alternative construction of eigenvectors. The following set of vectors $\ket{v_k}$ allows us to construct a code
\beq
\left(
  \begin{array}{c}
    \ket{v_k} \\
    \ket{v_{k+1}} \\
    \ket{v_{k+2}} \\
  \end{array}
\right):=\frac{1}{\sqrt{3}}
\left(
  \begin{array}{ccc}
    1 & \sqrt{2}\mathrm{e}^{\uroj \alpha_k} & 0 \\
    1 & -\frac{\sqrt{2}}{2}\mathrm{e}^{\uroj \alpha_k} & \frac{\sqrt{6}}{2}\mathrm{e}^{\uroj \alpha_k} \\
    1 & -\frac{\sqrt{2}}{2}\mathrm{e}^{\uroj \alpha_k} & -\frac{\sqrt{6}}{2}\mathrm{e}^{\uroj \alpha_k} \\
  \end{array}
\right)
\left(
  \begin{array}{c}
    \ket{u_k} \\
    \ket{u_{k+1}} \\
    \ket{u_{k+2}} \\
  \end{array}
\right),\quad k=1,4,7,10
\eeq
with some arbitrary phases $\alpha_j$ and $\ket{u_k}$ defined as previously.

One can check that in the considered case of highly regular spectrum requirement about non--degeneracy of spectrum can be relaxed and the phases $\xi_k$ for $k=5,6,7,8$ can be completely arbitrary.

Consider now a similar example in $d=3$ in which, instead of triangles, we will consider construction of sections crossing in the $z=0$ point, so that it belongs to the numerical range.

\begin{figure} [htbp]
\centerline{\epsfig{file=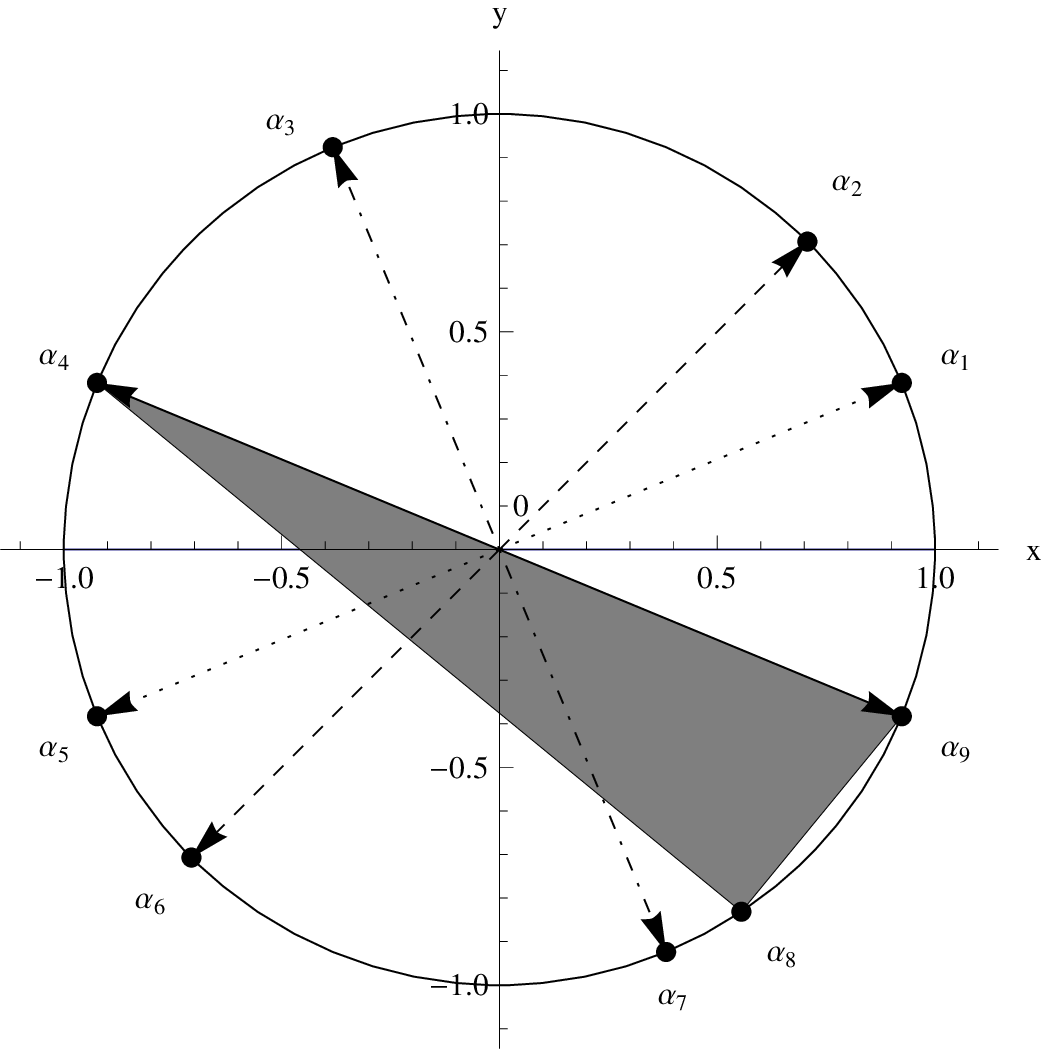, width=8.2cm}} 
\vspace*{13pt}
\fcaption{\label{motion} Exemplary distribution of eigenvalues on a unit circle. We have assumed, as further in the main text, that $\alpha_4=\alpha_9-\pi$. Moreover, $\alpha_1=\alpha_5-\pi$, $\alpha_2=\alpha_6-\pi$, $\alpha_3=\alpha_7-\pi$. The arrows represent the summation, Eq. (\ref{sum2}), giving rise to the compression value $0$.}\label{toy2}
\end{figure}

We have $U=\sum_{j=1}^9\eksp^{\uroj \alpha_j}\proj{\phi_j}$ with $\alpha_i > \alpha_j$ for $i>j$, and $\alpha_1=\alpha_5-\pi$, $\alpha_2=\alpha_6-\pi$, $\alpha_3=\alpha_7-\pi$ with an additional constraint that $0\in conv(\eksp^{\uroj \alpha_4}, \eksp^{\uroj \alpha_5}, \eksp^{\uroj \alpha_9})$. Fig. \ref{toy2} shows an exemplary distribution of eigenvalues.

In general:
\beq
R\otimes R'= U_A \left(
                   \begin{array}{ccc}
                     1 & 0 & 0 \\
                     0 & 1 & 0 \\
                     0 & 0& 0 \\
                   \end{array}
                 \right)
U_A^{\dagger}\otimes U_B \left(
                   \begin{array}{ccc}
                     1 & 0 & 0 \\
                     0 & 1 & 0 \\
                     0 & 0& 0 \\
                   \end{array}
                 \right)
U_B^{\dagger}=(\jedynka-\proj{f_3})\otimes (\jedynka-\proj{w_3})
\eeq
with some unitary $U_A$ and $U_B$ and corresponding vectors $\ket{f_3}$ and $\ket{w_3}$.

Set now the Fourier matrix $F_3$ in place of the local unitaries $U_A$ and $U_B$:
\beq
U_A=U_B=F_3=\frac{1}{\sqrt{3}}\left(
                                \begin{array}{ccc}
                                  1 & \omega & \omega^2 \\
                                  \omega^2 & 1 &\omega \\
                                  \omega & \omega^2 & 1 \\
                                \end{array}
                              \right),\quad \omega=\eksp^{\uroj \frac{2\pi}{3}}.
\eeq
Then $\ket{f_1}=(1,\omega,\omega^2)^T$, $\ket{f_2}=(\omega^2,1,\omega)^T$, $\ket{f_3}=(\omega,\omega^2,1)^T$, $\ket{w_i}=\ket{f_i}$. We now define properly eigenstates of $U$, which could lead to a product code $P=\sum_{i=1}^4 \proj{\psi_i}= R\otimes R'$, $\ket{\psi_1}=\ket{f_1}\otimes\ket{f_1}$, $\ket{\psi_2}=\ket{f_1}\otimes\ket{f_2}$, $\ket{\psi_3}=\ket{f_2}\otimes\ket{f_1}$, $\ket{\psi_4}=\ket{f_2}\otimes\ket{f_2}$, with a simplifying assumption $\alpha_4=\alpha_9-\pi$.  By choosing our eigenvalues so that
\beqn\label{sum2}
\frac{1}{2} z_{j}+\frac{1}{2} z_{j+5}=0,\quad j=1,2,3,4
\eeqn
we ensured that we can set
\beqn
\ket{\psi_j}:=\frac{1}{\sqrt{2}} \left(\ket{\phi_{j}}+\ket{\phi_{j+5}}\right),\quad j=1,2,3,4.
\eeqn
It remains to choose $\ket{\phi_i}$ so that $\ket{\psi_i}$ are product. For example, one choice could be
\beqn
\ket{\phi_{1,6}}=\frac{1}{\sqrt{2}}(\ket{f_1}\otimes\ket{f_1}\pm\ket{f_3}\otimes\ket{f_2}),\quad
\ket{\phi_{2,7}}=\frac{1}{\sqrt{2}}(\ket{f_1}\otimes\ket{f_2}\pm\ket{f_2}\otimes\ket{f_3}),\\
\ket{\phi_{3,8}}=\frac{1}{\sqrt{2}}(\ket{f_2}\otimes\ket{f_1}\pm\ket{f_1}\otimes\ket{f_3}),\quad
\ket{\phi_{4,9}}=\frac{1}{\sqrt{2}}(\ket{f_2}\otimes\ket{f_2}\pm\ket{f_3}\otimes\ket{f_1}).
\eeqn

Our construction is in fact such that $0$ belongs to the symmetric product higher rank numerical range.

\section{Discussion and conclusions}\label{dyskusja}

Motivated by the form of Knill--Laflamme conditions for multiple access channels, we have introduced the notion of the  product higher rank numerical range as a tool helpful in constructing quantum error correction codes for such type of quantum channels. Several useful extensions of it, namely: the symmetric product range and the common product range have also been discussed. Techniques for bounding the product range and some analytical techniques for findings ones in some cases have been introduced.
We have applied our findings to a construction of error correction codes for a class of two--access biunitary quantum channel.
The reverse problem of finding the noise model for a given product error correction code has also been considered.

Concluding, we state some open problems related to the subject. Among them, determination of the shape of the product higher rank numerical range for different types of operators seems to be one of the most important ones. In particular, for the case of $d=3$, it should be verified whether the $2\otimes 2$ product range of a projection operator is at most a one--element set. If this is indeed the case, the uniqueness (Observation \ref{only}) is a general feature in this setting.
Also the issue of simple-connectivity of the set $\kan_{k_1\otimes k_2\otimes \cdots}(A)$ in arbitrary dimensions is of particular interest. From the point of view of quantum error correction special attention should be devoted to normal operators.

It is worth adding that the issue of whether a given value $\tilde{\lambda}$ may belong to
the product range of a given operator
 (and determining the corresponding projection if it does belong) is
closely related to the problem of
estimating  local norms  of operators acting on composite Hilbert
spaces \cite{normy-kribs}.
For concreteness consider an operator $A$ acting on a bipartite Hilbert
space $\calH_1 \otimes \calH_2$
and a  projector  $P_0=\sum_i \proj{i}$. To find the local norm of $A-\tilde{\lambda}\jedynka$
one needs to
    find the following minimum over local unitaries
$D:= \min_{U,V}    || (P_0\otimes P_0) (U\otimes V) A (U^{\dagger}\otimes V^{\dagger})
(P_0\otimes P_0)    -\tilde{\lambda} P_0\otimes P_0||$.
Clearly, if D cannot be made smaller than the prescribed accuracy then
$\tilde{\lambda}$ cannot belong to
the product range. It is thus important to design efficient procedures for
computing local norms
of operators acting on composite Hilbert spaces and
 finding explicit forms of product projections from the product codes set.

\nonumsection{Acknowledgements} \noindent MD was supported by
Gda\'nsk University of Technology through the grant "Dotacja na
kszta\l cenie m\l odych kadr w roku 2011". PH is supported by
Polish Ministry of Science under Grant No. NN202231937. He also
acknowledges partial support from  ERC under Advanced Grant
QOLAPS. K\.Z is supported by National Science Centre through the
project Maestro DEC-2011/02/A/ST2/00305.

\nonumsection{References}

\end{document}